\documentclass[11pt]{article}
\usepackage{moriond}

\def\be{\begin{equation}}
\def\ee{\end{equation}}
\def\bea{\begin{eqnarray}}
\def\eea{\end{eqnarray}}

\newcommand{\Photo}{\includegraphics[height=35mm]{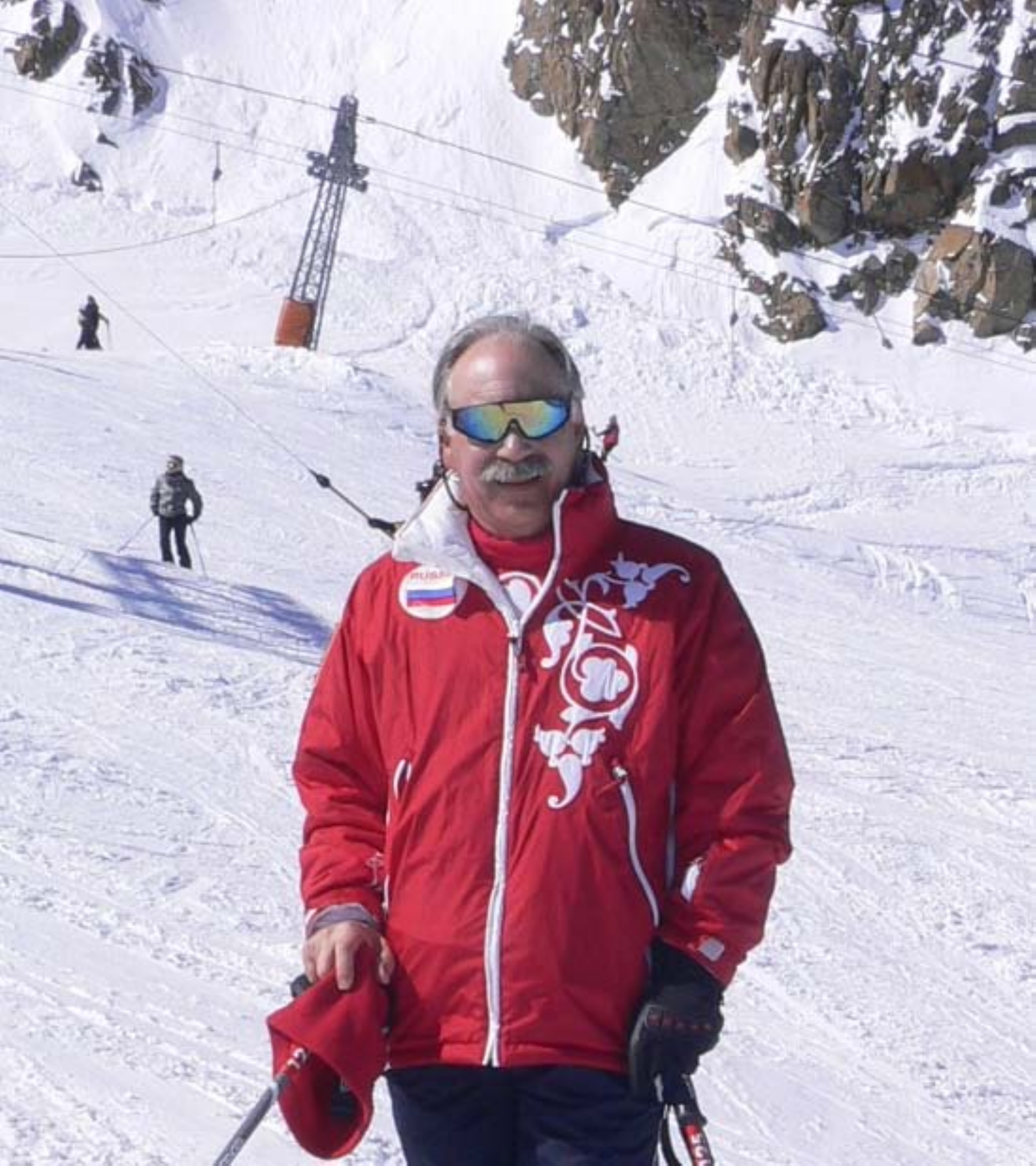}}

\begin{document}
\vspace*{4cm}
\title{SUSY PHENOMENOLOGY TODAY \footnote{Talk at XLVIIIth Rencontres de Moriond  "Electroweak interactions and Unified theories", March 7th, 2013. }}

\author{ D.I.KAZAKOV}

\address{Bogoliubov Laboratory of Theoretical Physics, Joint Institute for Nuclear Research, \\141 980, Dubna, Moscow region, Russia}

\maketitle\abstracts{
The review of the SUSY phenomenology today is given with the emphasis  on generic properties of SUSY models,
SUSY searches at the LHC and in astrophysics. Due to the absence of any definite SUSY signal at accelerators and other experiments, we describe the allowed regions in the MSSM and NMSSM parameter space and discuss the possibilities of SUSY manifestation in the near future.}

\section{Introduction. Why SUSY?}

First of all, let me remind you why we continue to explore SYSY despite the absence of any
of its manifestations  in experiments.\\
SUSY at TeV scale:

$\bullet$ Provides unification of the gauge couplings

$\bullet$ Solves the hierarchy problem

$\bullet$ Explains the electroweak symmetry breaking

$\bullet$ Provides the Dark matter candidate\\
SUSY in particle physics:

$\bullet$ Provides unification with gravity

$\bullet$ Required for the String/Brane picture

$\bullet$ Maximal SUSY theories might be integrable -
a way to non-perturbative solutions and quantum gravity.

It should be noted that while an application of supersymmetry to particle physics is still questionable, mathematical supersymmetric theories  exist and promise many new discoveries due to their remarkable properties.  Here, I am going to concentrate on the SUSY phenomenology  having in mind particle physics and partially astrophysics.

To be more precise, one has to specify which SUSY model is discussed. There are many models that pretend to describe the phenomenology of particle physics. They are all based on the so-called N=1 supersymmetry and generalize the Standard Model of three fundamental interactions but differ in a way how supersymmetry is broken. The incomplete list of SUSY models consists of:
MSSM, \ CMSSM, \ mSUGRA, \ mGMSB, \  mAMSB, \  NUHM, \  NMSSM, \  No Scale, \  Split SUSY, \ pMSSM, \  etc.

Despite supersymmetric rigidity of dimensionless couplings, the arbitrariness of soft terms makes 
predictions strongly model dependent! In particular, the mass spectrum of super partners varies from model to model and  in principle allows one to distinguish between them.
The particle content of all these models is the same with possible additional multiplets like in the NMSSM. In all versions of the MSSM the particle content is
\begin{center}
\vglue 0.6cm
\nopagebreak[4]
\renewcommand{\tabcolsep}{0.03cm}
\begin{tabular}{lllccc}
Superfield & \ \ \ \ \ \ \ Bosons & \ \ \ \ \ \ \ Fermions &
$SU_c(3)$& $SU_L(2)$ & $U_Y(1)$ \\ \hline \hline Gauge  &&&&& \\
${\bf G^a}$   & gluon \ \ \ \ \ \ \ \ \ \ \ \ \ \ \  $g^a$ &
gluino$ \ \ \ \ \ \ \ \ \ \ \ \ \tilde{g}^a$ & 8 & 1 & 0 \\ ${\bf
V^k}$ & Weak \ \ \ $W^k$ \ $(W^\pm, Z)$ & wino, zino \
$\tilde{w}^k$ \ $(\tilde{w}^\pm, \tilde{z})$ & 1 & 3& 0 \\ ${\bf
V'}$   & Hypercharge  \ \ \ $B\ (\gamma)$ & bino \ \ \ \ \ \ \ \ \
\ \ $\tilde{b}(\tilde{\gamma })$ & 1 & 1& 0 \\ \hline Matter &&&&
\\ $\begin{array}{c} {\bf L_i} \\ {\bf E_i}\end{array}$ & sleptons
\ $\left\{
\begin{array}{l} \tilde{L}_i=(\tilde{\nu},\tilde e)_L \\ \tilde{E}_i =\tilde
e_R \end{array} \right. $ & leptons \ $\left\{ \begin{array}{l}
L_i=(\nu,e)_L
\\ E_i=e_R \end{array} \right.$ & $\begin{array}{l} 1 \\ 1 \end{array} $  &
$\begin{array}{l} 2 \\ 1 \end{array} $ & $\begin{array}{r} -1 \\ 2
\end{array} $ \\ $\begin{array}{c} {\bf Q_i} \\ {\bf U_i} \\ {\bf D_i}
\end{array}$ & squarks \ $\left\{ \begin{array}{l}
\tilde{Q}_i=(\tilde{u},\tilde d)_L \\ \tilde{U}_i =\tilde u_R \\
\tilde{D}_i =\tilde d_R\end{array}  \right. $ & quarks \ $\left\{
\begin{array}{l} Q_i=(u,d)_L \\ U_i=u_R^c \\ D_i=d_R^c \end{array}
\right.$ & $\begin{array}{l} 3
\\ 3^* \\ 3^* \end{array} $  & $\begin{array}{l} 2 \\ 1 \\ 1 \end{array} $ &
$\begin{array}{r} 1/3 \\ -4/3 \\ 2/3 \end{array} $ \\ \hline Higgs
&&&& \\ $\begin{array}{c} {\bf H_1} \\ {\bf H_2}\end{array}$ &
Higgses \ $\left\{
\begin{array}{l} H_1 \\ H_2 \end{array}  \right. $ & higgsinos \ $\left\{
 \begin{array}{l} \tilde{H}_1 \\ \tilde{H}_2 \end{array} \right.$ &
$\begin{array}{l} 1 \\ 1 \end{array} $  & $\begin{array}{l} 2 \\ 2
\end{array} $ &
$\begin{array}{r} -1 \\ 1
\end{array} $
 \\ \hline \hline
\end{tabular}
\end{center}\vspace{0.3cm}
One has to underline that all particles of the SM have super partners and belong to either chiral or vector representation of the SUSY algebra and none of the SM particles is a super partner to another.

\section{Manifestation of SUSY}
Search for SUSY manifestation in Nature has a long but unsuccessful history. In particle physics it includes:

$\bullet$ Direct production at colliders at high energies 

$\bullet$ Indirect manifestation at low energies

-- Rare decays ( $B_s\to s\gamma, B_s\to \mu^+\mu^- , B_s\to \tau\nu $) 

-- g-2 of the muon

$\bullet$ Search for long-lived SUSY particles \\
In astrophysics (provided SUSY composes the DM):

$\bullet$ Relic abundancy of Dark Matter in the Universe

$\bullet$ DM annihilation signal in cosmic rays 

$\bullet$ Direct DM interaction with nucleons

Nothing has been found so far....

\begin{figure}
\begin{center}
\includegraphics[height=70mm]{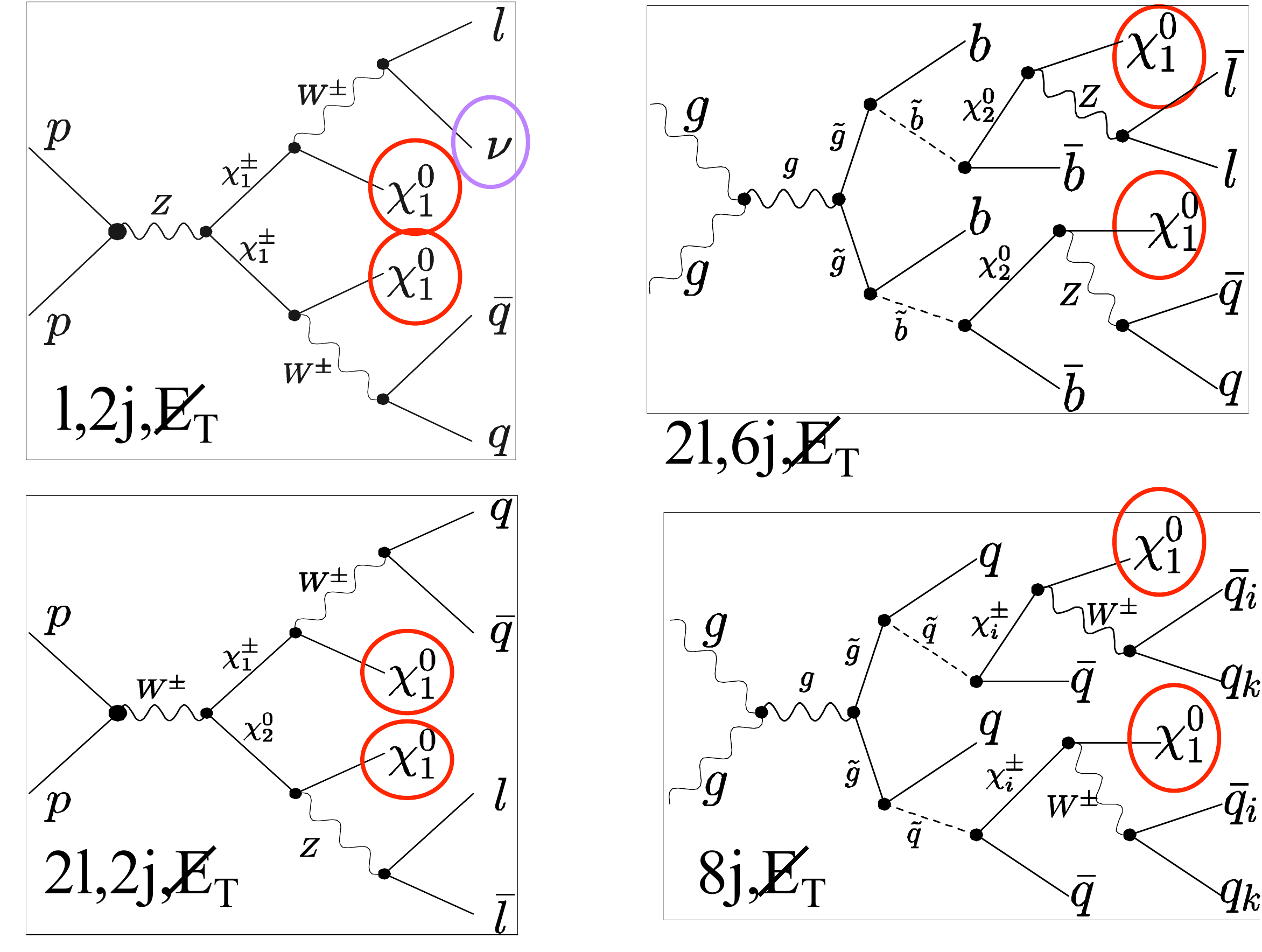}
\caption{Typical processes of super partner creation and cascade decay in the MSSM}\label{1}
\end{center}
\end{figure}

At the LHC  we are looking for creation  and decay of superpartners in cascade processes. The typical processes are shown in Fig.\ref{1}.
The typical SUSY signature is the missing energy and transverse momentum carried by a lightest super partner.

There are two ways of presenting and analyzing data on SUSY searches:

1. \underline{High energy input}: introduce the universal parameters at high energy scale (GUT). For example, $m_0, m_{1/2}, A_0, 
\tan\beta$ in the MSSM. Then run them down to low energies using RG equations.
The advantage is a small number of universal parameters for all masses. The disadvantage is the strict model dependence (MSSM, NMSSM, etc).

2. \underline{Low energy input}: use the low energy parameters like masses of super partners, mixings, etc. For 
example, $\tilde m_g, \tilde m_q, \tilde m_\chi$ or $m_A, \tan\beta$.
The advantage is the less model dependence. The
disadvantage is the increase of parameters and process dependence.

Both the approaches are used in analyses.

\section{The progress of the LHC}

In the proton-proton collisions at the LHC the supersymmetric
particles can be produced according to the diagrams like those shown
in  Fig.~\ref{1}. 
The ``strong'' production cross sections are characterized by
a large number of jets from the long decay chains and the
missing energy from the escaping neutralino. These
characteristics can be used to efficiently suppress the
background. For the electroweak production, both the number
of jets and the missing transverse energy are low since the
LSP is not boosted so strongly as in the decay of  heavier
strongly interacting particles. Hence, the electroweak gaugino
production needs the leptonic decays to reduce the background,
so these signatures need more luminosity and cannot compete at
present with the sensitivity of the ``strong'' production of the
squarks and gluinos.

\begin{figure}[htb]
\begin{center}
\leavevmode
\includegraphics[width=0.45\textwidth,height=4.9cm]{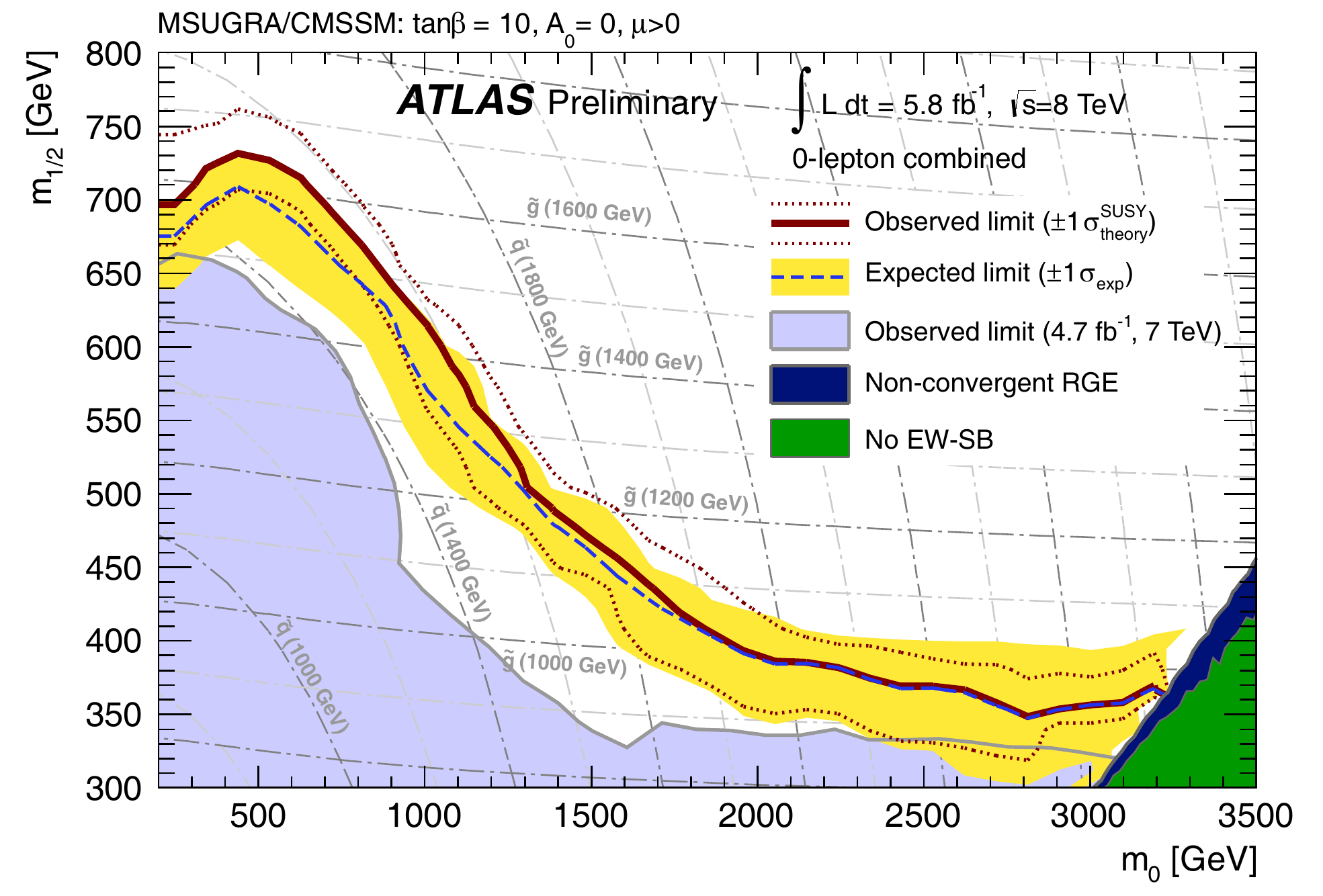}
\includegraphics[width=0.45\textwidth,,height=4.9cm]{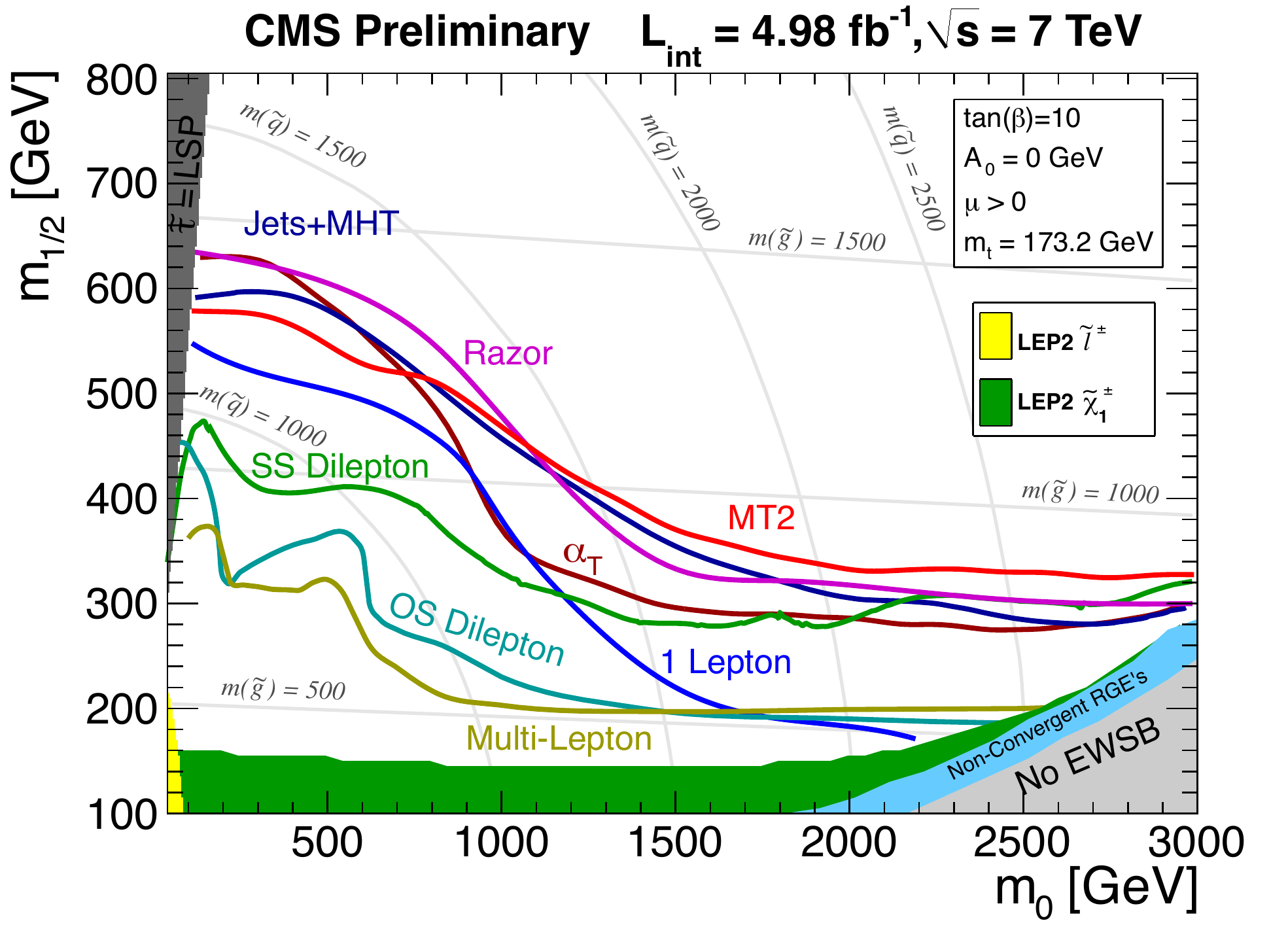}
\end{center}
 \caption{SUSY searches presented as exclusion plots in the $m_0 - m_{1/2}$ plane}
\label{2}
\end{figure}

We present here some examples of superparticle searches in
various scenarios depicted as exclusion plots. Everywhere in
these plots the excluded region is the one below the
corresponding curve (lower masses, lower values of parameters).
We first demonstrate the results obtained along the first (or high energy) approach.
They are typically presented in the $m_0 - m_{1/2}$ plane keeping the other parameters either fixed or adjusted
at each point (see Fig.\ref{2}). 
These constraints on the values of $m_0,m_{1/2}$ are equally valid for any process,
but refer to the CMSSM framework \cite{ATLAS_SUSY},\cite{CMS_SUSY_pub}. 

\begin{figure}[htb]
\begin{center}
\includegraphics[width=0.45\textwidth,height=5cm]{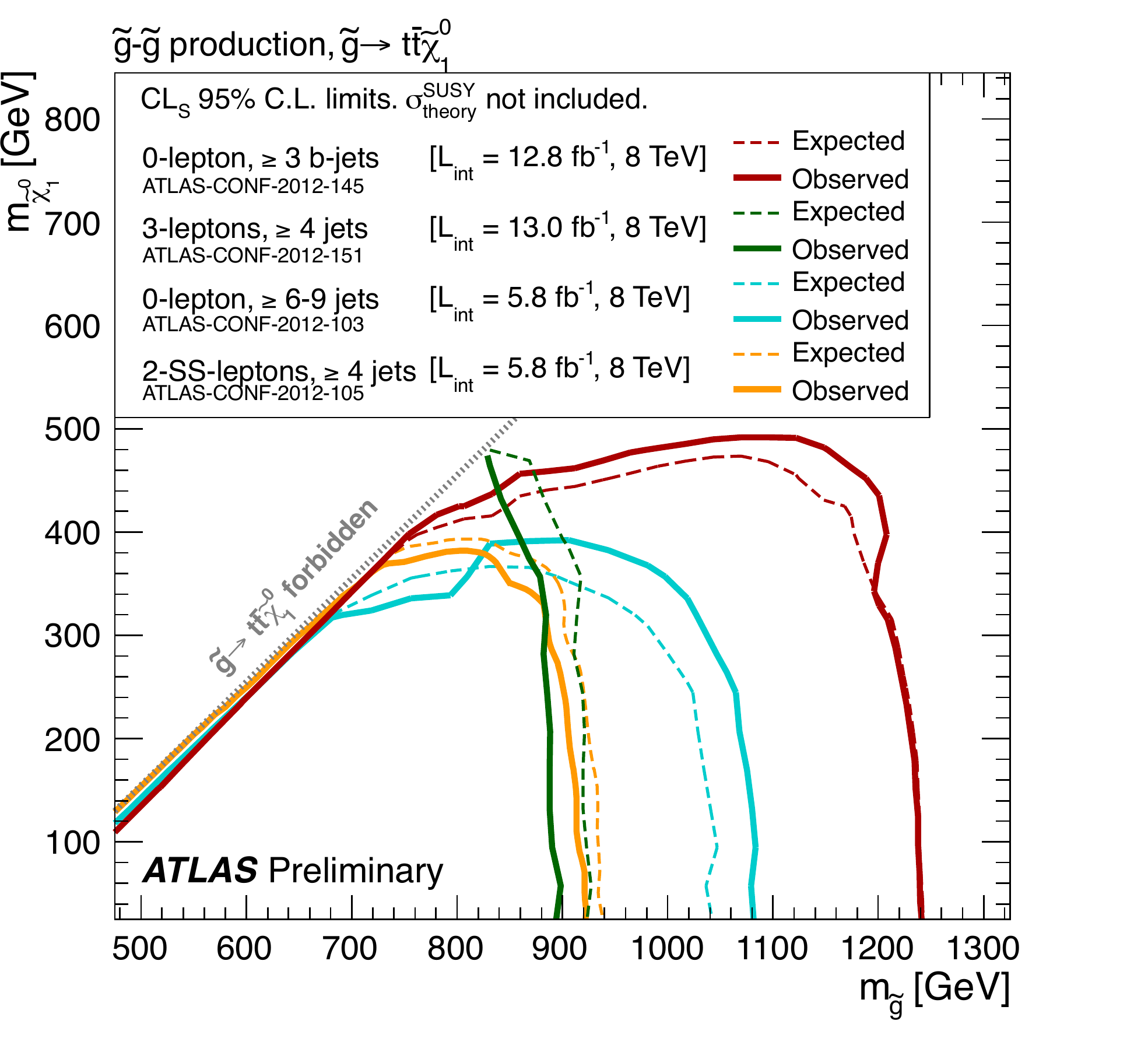}
\includegraphics[width=0.45\textwidth,height=5cm]{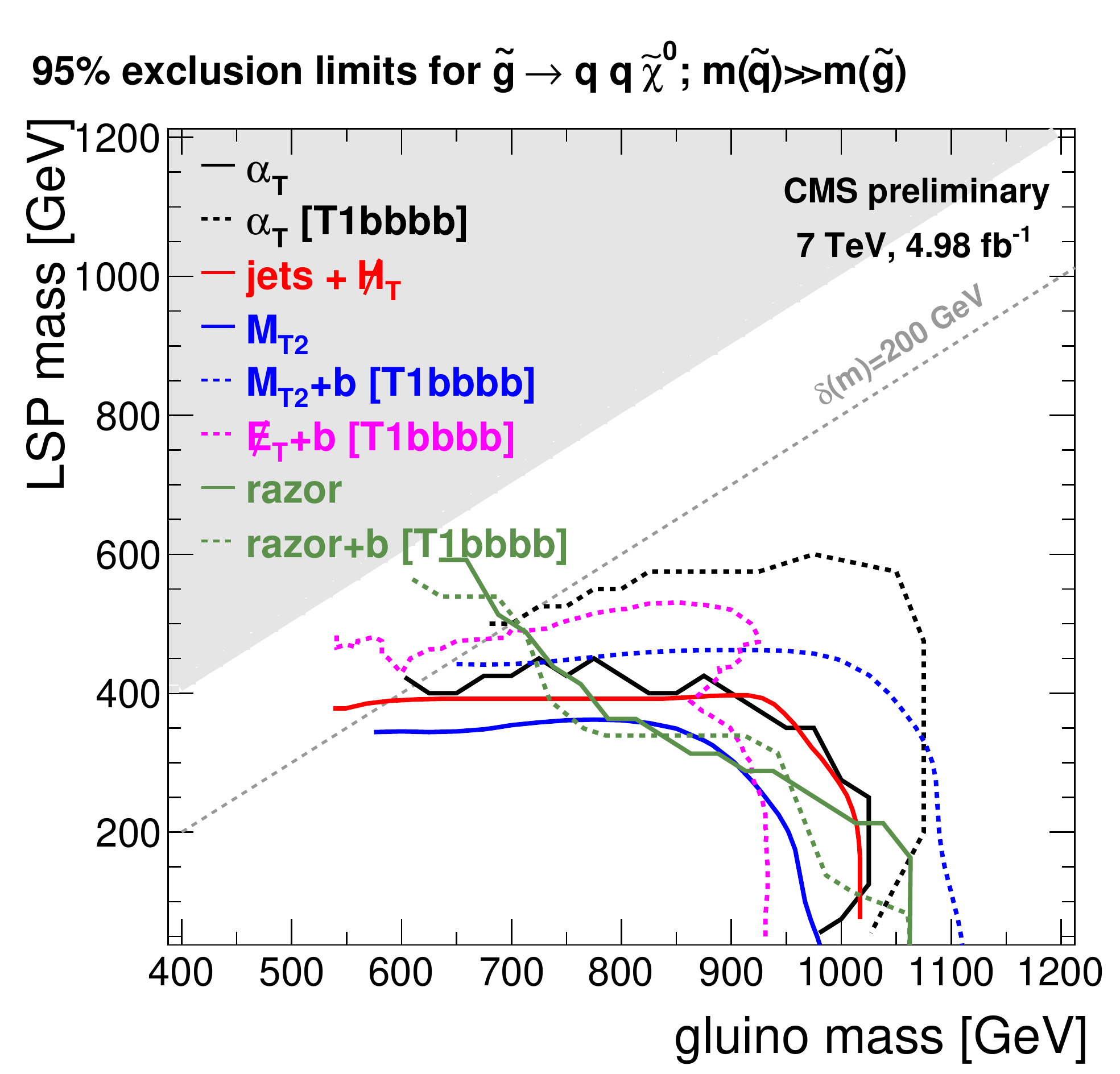}
\includegraphics[width=0.45\textwidth,height=5cm]{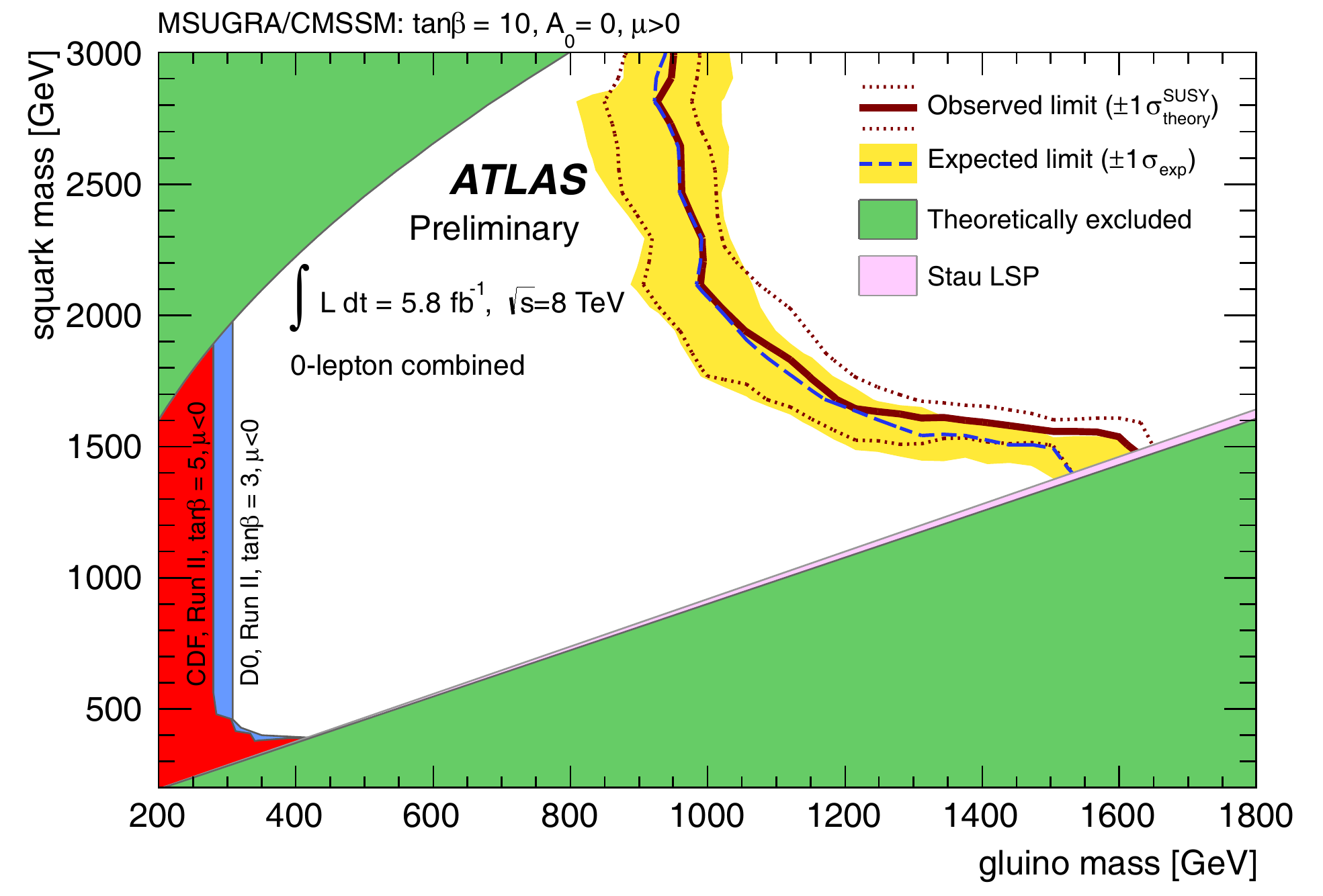}
\includegraphics[width=0.45\textwidth,height=5cm]{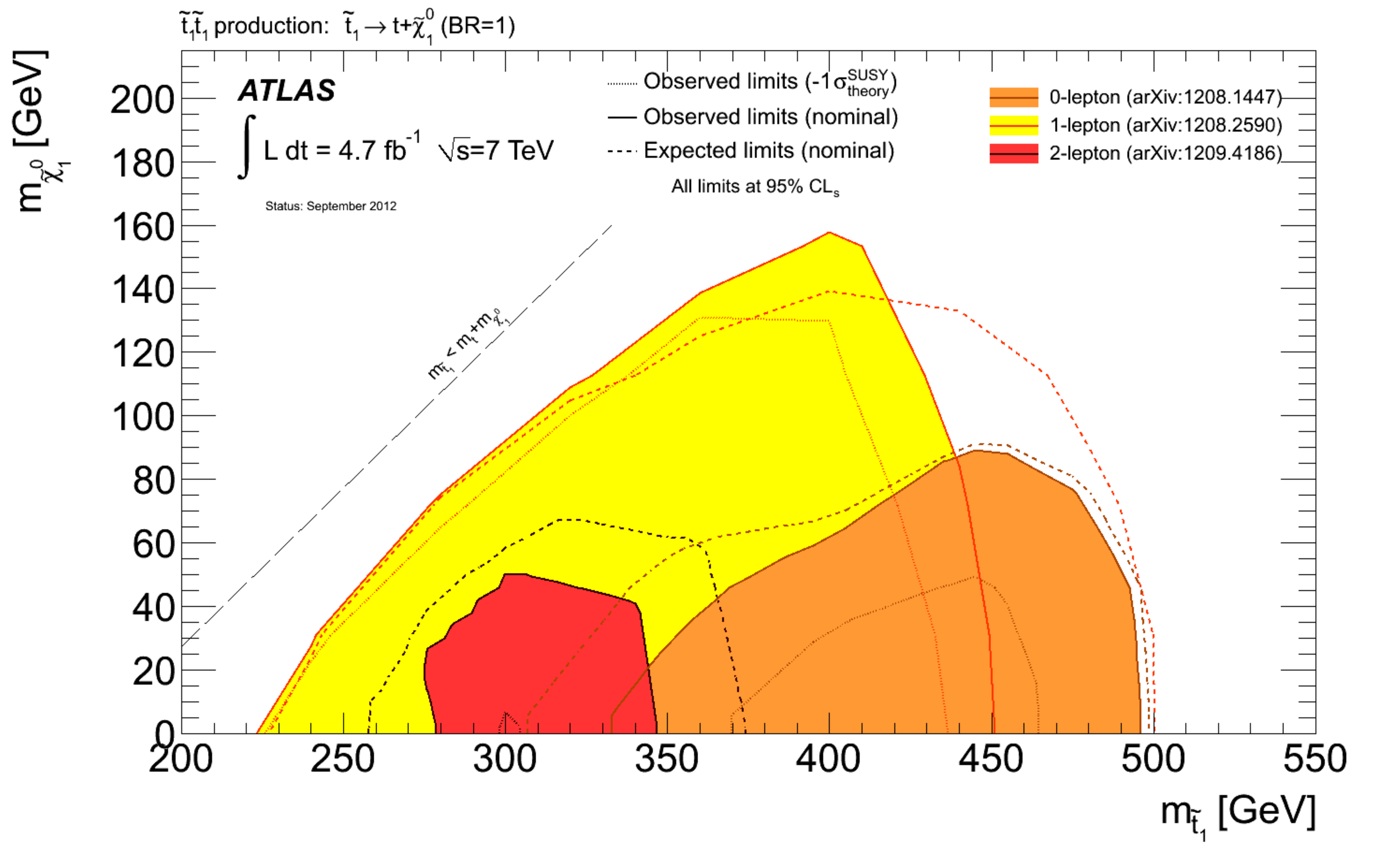}
\end{center}
\caption{Upper row:
95\% CL exclusion limits for MSUGRA/CMSSM models with
$\tan\beta = 10$, $A_0 = 0$ and $mu > 0$;
Left: ATLAS exclusion limits at 95\%~CL for 8~TeV
analyses in the $m_{\tilde g}-m_{\tilde\chi^0_1}$ plane for the
\emph{Gtt} simplified model where a pair of gluino decays via
off-shell stop to four top quarks and two neutralinos (LSP).
Right: CMS exclusion limits at 95\%~CL for 7~TeV analyses of
gluino decay $\tilde g \to qq \tilde\chi^0_1$, assuming
$m_{\tilde q} >> m_{\tilde g}$.}
\label{3}
\end{figure}

The second (low energy) approach corresponds to particular processes, as shown in Fig.\ref{3}. 
Direct searches for  super partners at the LHC in different
channels have pushed the lower limits on their masses, mainly
of the gluinos and the squarks of the first two generations,
upwards to the TeV range. 
The fisrt example is the gluino pair production
$pp \to \tilde g \tilde g$ and
$\tilde g \to t \bar t \tilde\chi^0_1$ decay.
Four different final states (0 leptons
with $\ge$ 3 $b$-jets; 3 leptons with
$\ge$ 4 jets; 0 leptons with $\ge$ 6-9
jets; and a pair of the same-sign
leptons with more than 4 jets) are
considered \cite{ATLAS_SUSY},\cite{CMS_SUSY_pub}. The analysis was performed using
13.0~fb$^{-1}$ data results in the non-observation
of the gluino lighter than 900~GeV (conservative limit) or even
1200~GeV for the lightest neutralino mass less than around
300~GeV, and squarks  of the first two generations lighter than 1500 GeV.

\begin{figure}[htb]
\begin{center}
\leavevmode
\includegraphics[width=0.45\textwidth,height=6cm]{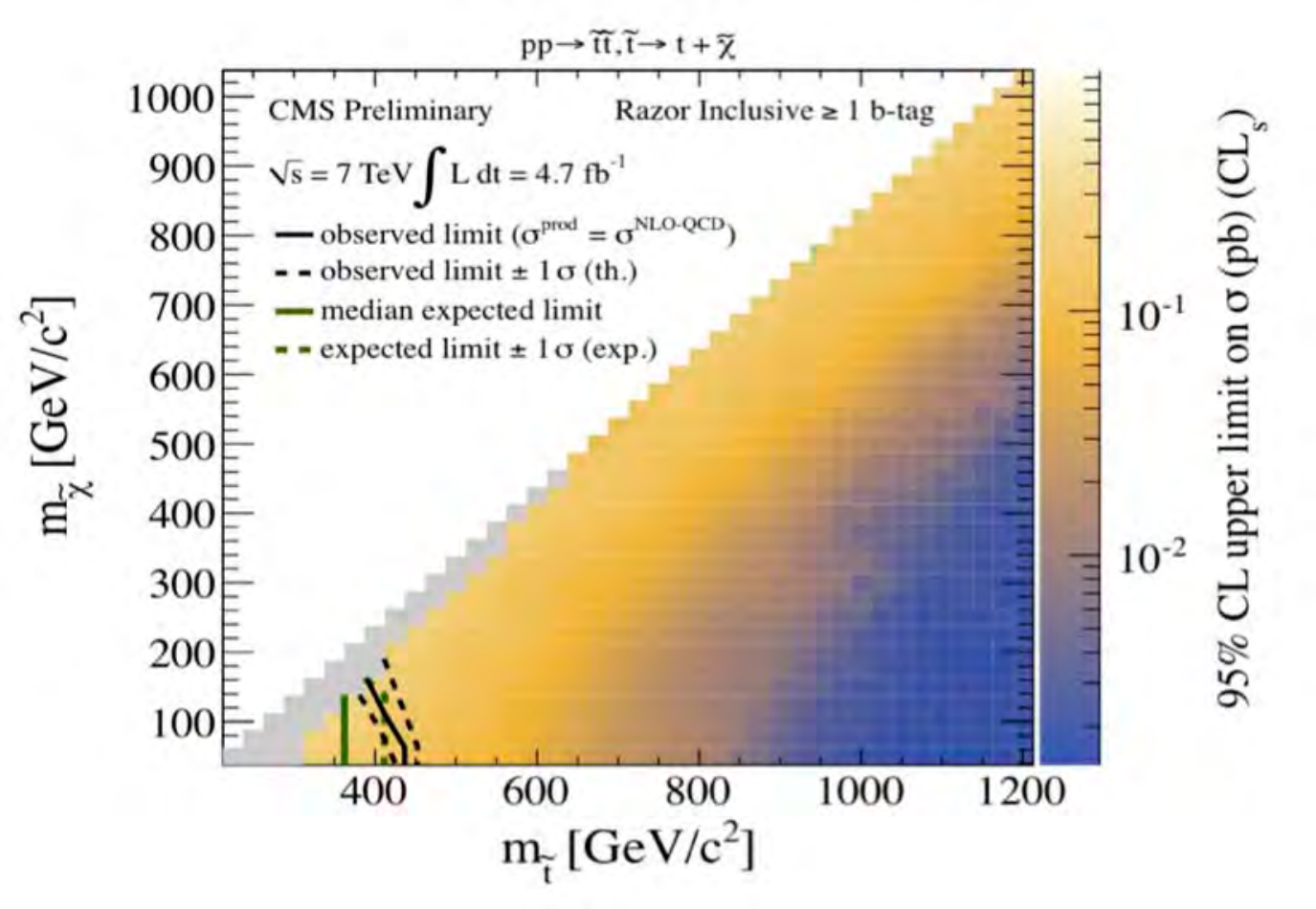}
\includegraphics[width=0.45\textwidth,height=6cm]{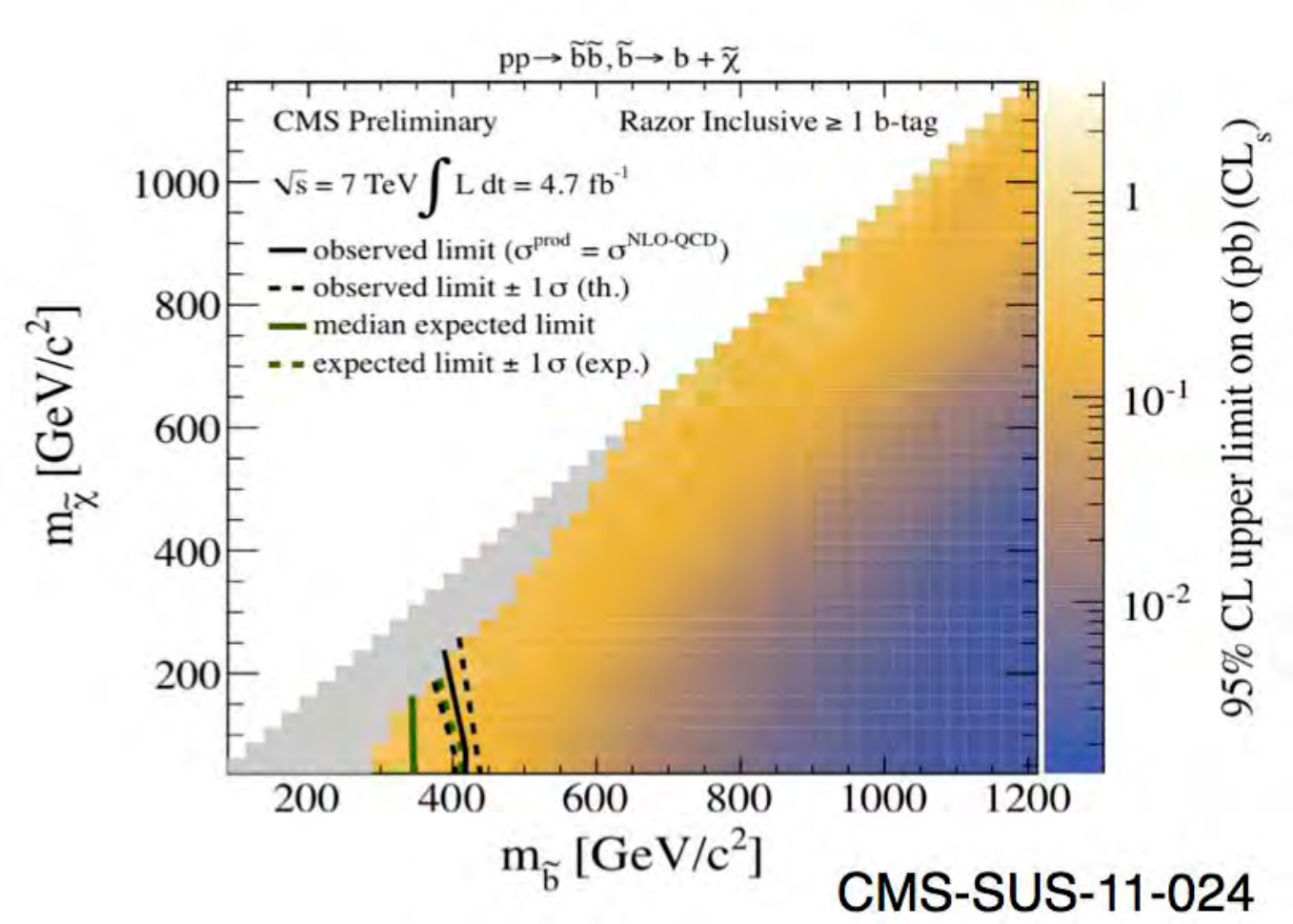}
\end{center}
\caption{Stop and bottom searches at the LHC}
\label{4}
\end{figure}

On the other hand, for the third
generation the limits are rather weak and the masses around 
a few hundred~GeV are still allowed, as can be seen in the right bottom panel of Fig.\ref{3} and Fig.\ref{4}. 
This is due to different decay modes originated from large Yukawa couplings of the third generation.
 The exclusion
limits at 95\%~CL are shown in the $\tilde t_1 - \tilde\chi^0_1$
mass plane. Depending on the stop mass there can be
two different decay channels. For relatively light stops with
masses below 200~GeV, the decay $\tilde t_1 \to b +
\tilde\chi^\pm_1$, $\tilde\chi^\pm_1 \to W^* + \tilde\chi^0_1$
is assumed with two hypotheses on the
$\tilde\chi^\pm_1$, $\tilde\chi^0_1$ mass hierarchy,
$m(\tilde\chi^\pm_1) = 106$~GeV and $m(\tilde\chi^\pm_1) =
2 m(\tilde\chi^0_1)$~\cite{ATLAS_stop1}.
For the heavy stop masses above 200~GeV, the decay $\tilde t_1
\to t + \tilde\chi^0_1$ is assumed to dominate.
\begin{figure}[htb]
\begin{center}
\leavevmode
\includegraphics[width=0.45\textwidth,height=5cm]{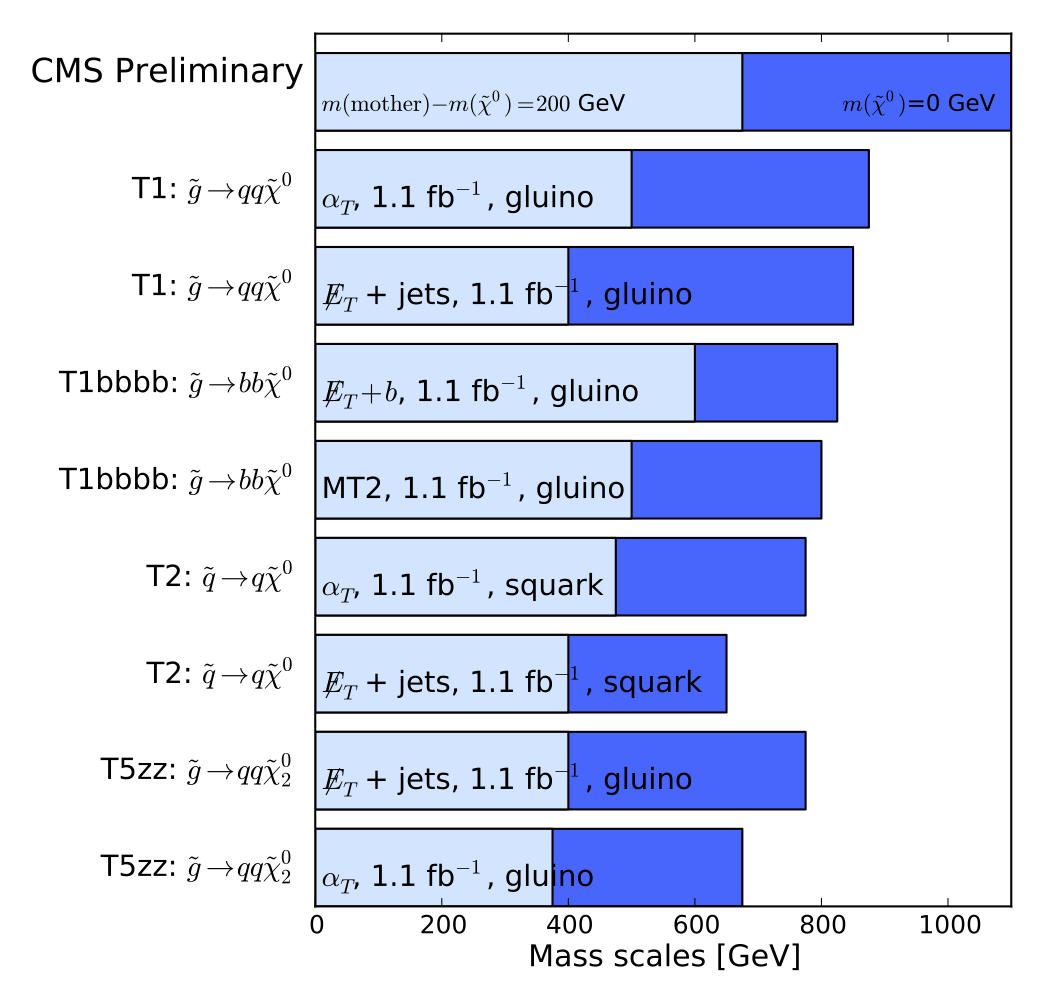}
\includegraphics[width=0.45\textwidth,height=5cm]{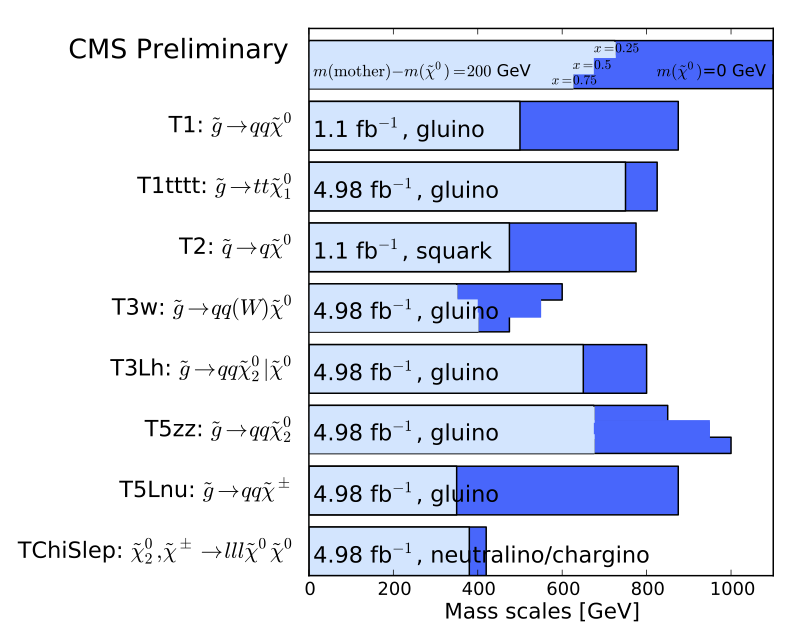}
\vspace*{-3mm}
\end{center}
\caption{Best exclusion limits for the gluino and squark
masses, for $m_{\chi^0}$ = 0~GeV (dark blue) and
$m(mother) - m_{\chi^0}$ = 200~GeV (light blue), for each
topology, for the hadronic results}
\label{5}
\end{figure}

All the exclusion plots discussed above can give direct limits
on the masses of supersymmetric particles under certain
assumptions (mass relations, dominant decay channels, modified
or/and simplified models, etc.). The latest mass limits for the
different models and final state channels obtained by the CMS collaboration are
shown in Fig.~\ref{5}~\cite{CMS_SUSY_pub,CMS_SUSY_web}.

\begin{figure}[htb]
\begin{center}
\includegraphics[width=0.48\textwidth]{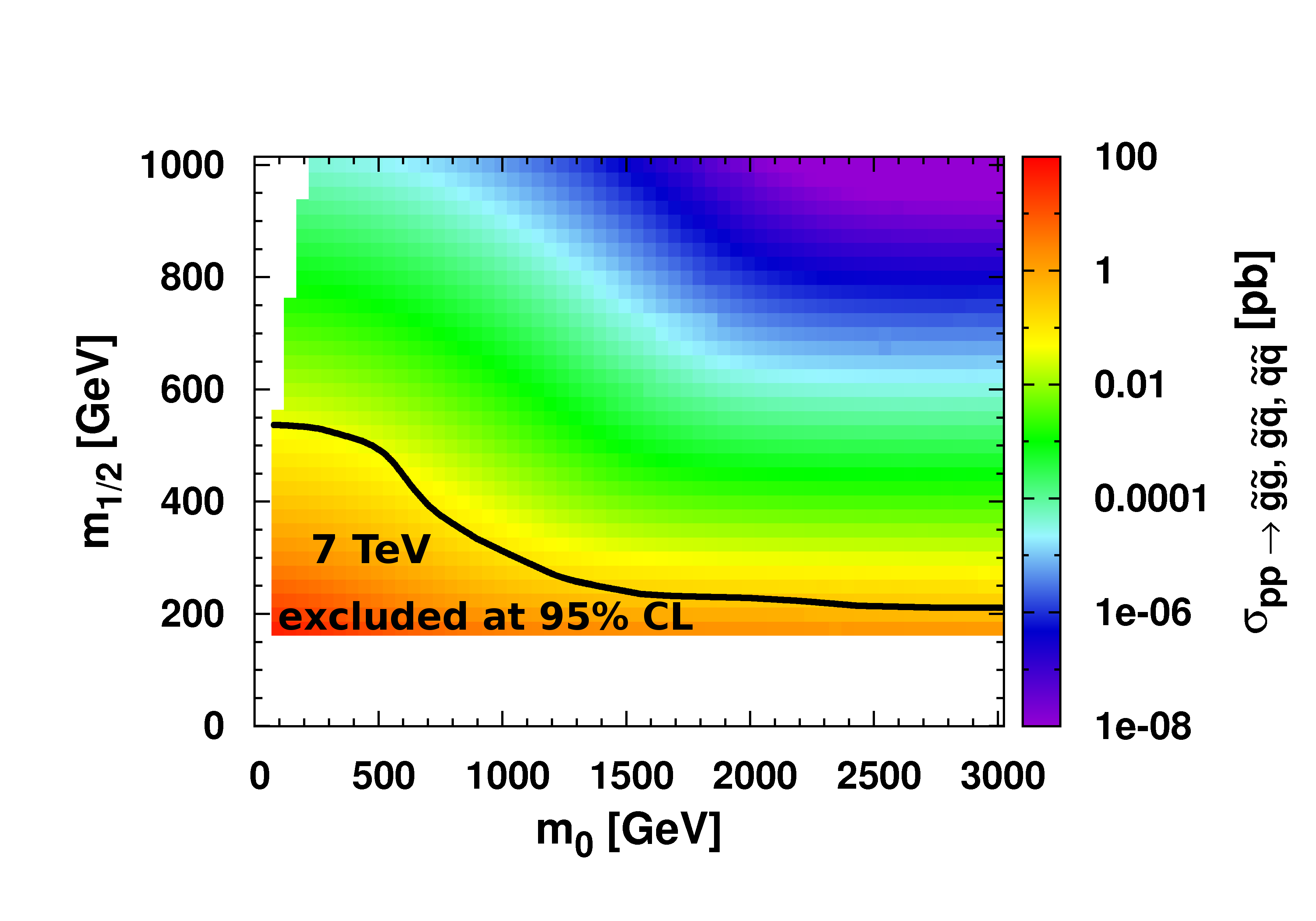}
\includegraphics[width=0.48\textwidth]{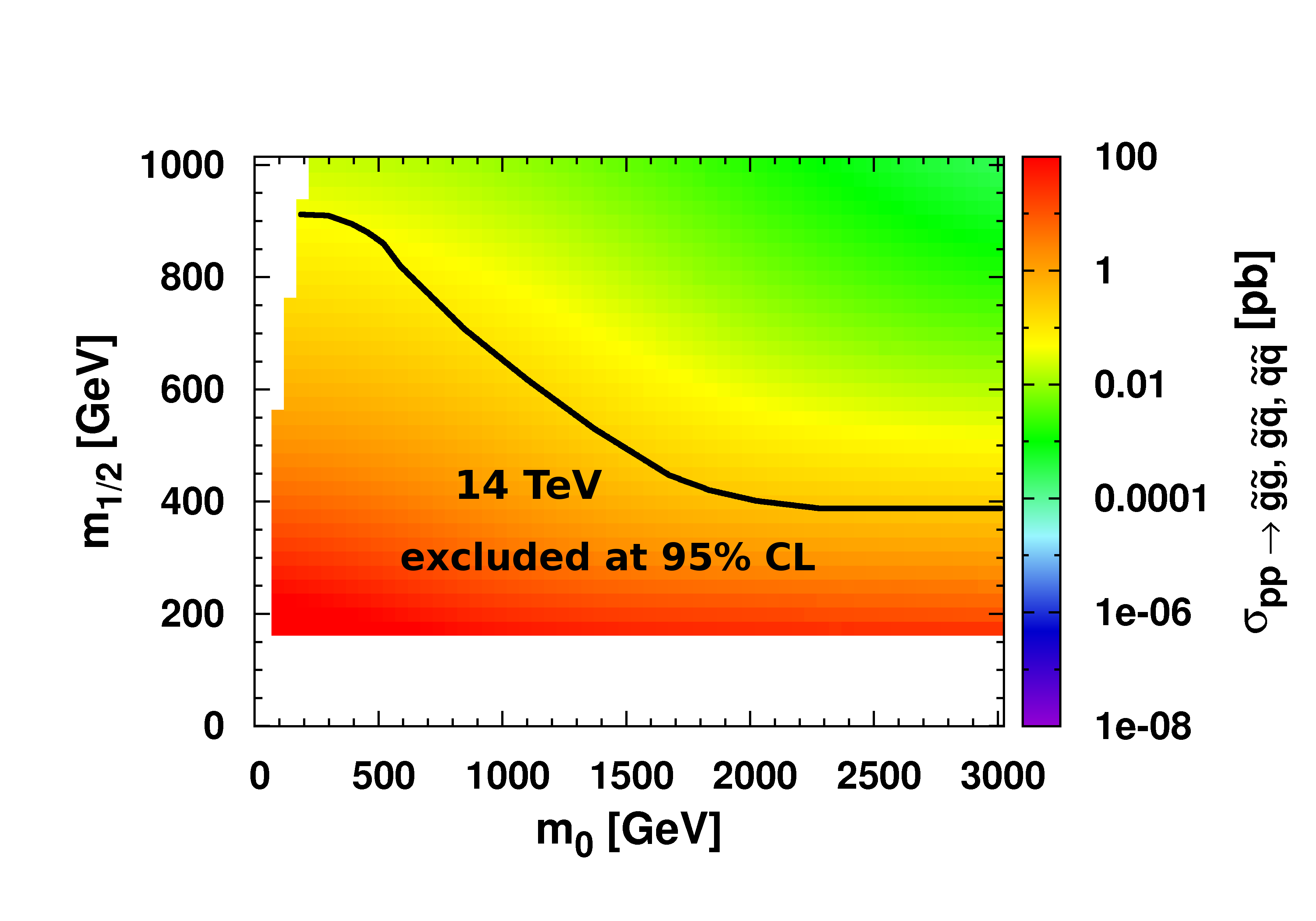}
\vspace{-5mm}
\end{center}
\caption{Left: The total production cross-section of the
strongly interacting particles in comparison with the LHC
excluded limits for 7+8~TeV. Here the data from ATLAS and CMS
were combined and correspond to the integrated luminosity of
1.3 and 1.1~fb$^{-1}$, respectively. One observes that the
cross-section of 0.1 to 0.2~pb is excluded at 95\% CL.
Right: the cross~sections at 14~TeV and expected
exclusion for the same limit on the cross-section as at 7~TeV.}
\label{f4}
\end{figure}

The total cross-section for strongly interacting particles
is shown in Fig.~\ref{f4} together with the excluded region
from the direct searches for SUSY particles at the LHC \cite{bbkr2}. One
observes that the excluded region (below the solid line) follows
rather closely the total cross-section indicated by the colour
shading. From the colour coding one observes that the excluded
region corresponds to the cross-section limit of about
$0.1-0.2$~pb.

\begin{figure}[htb]
\begin{center}
\includegraphics[width=0.46\textwidth]{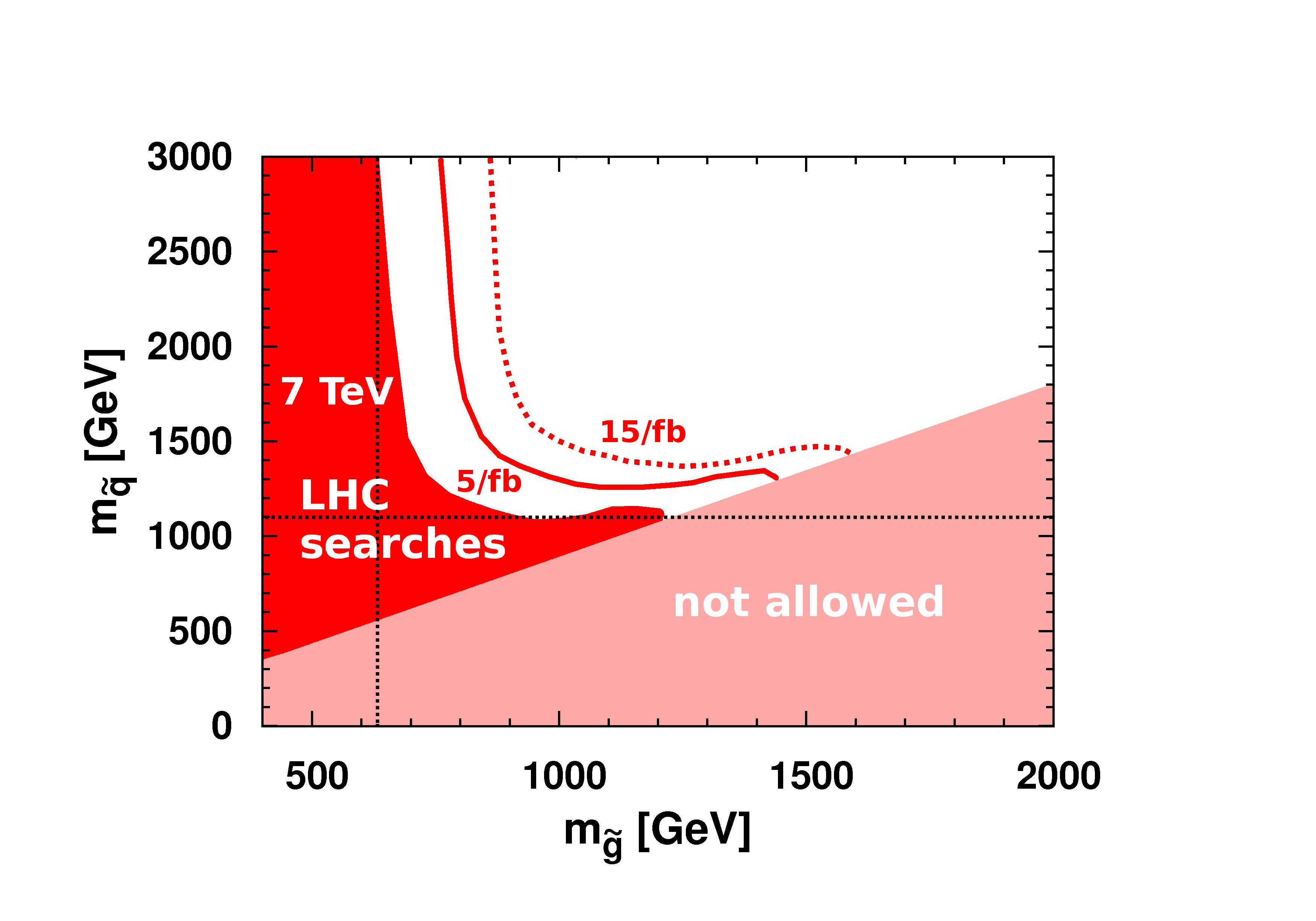}
\raisebox{-5pt}{\includegraphics[width=0.49\textwidth]{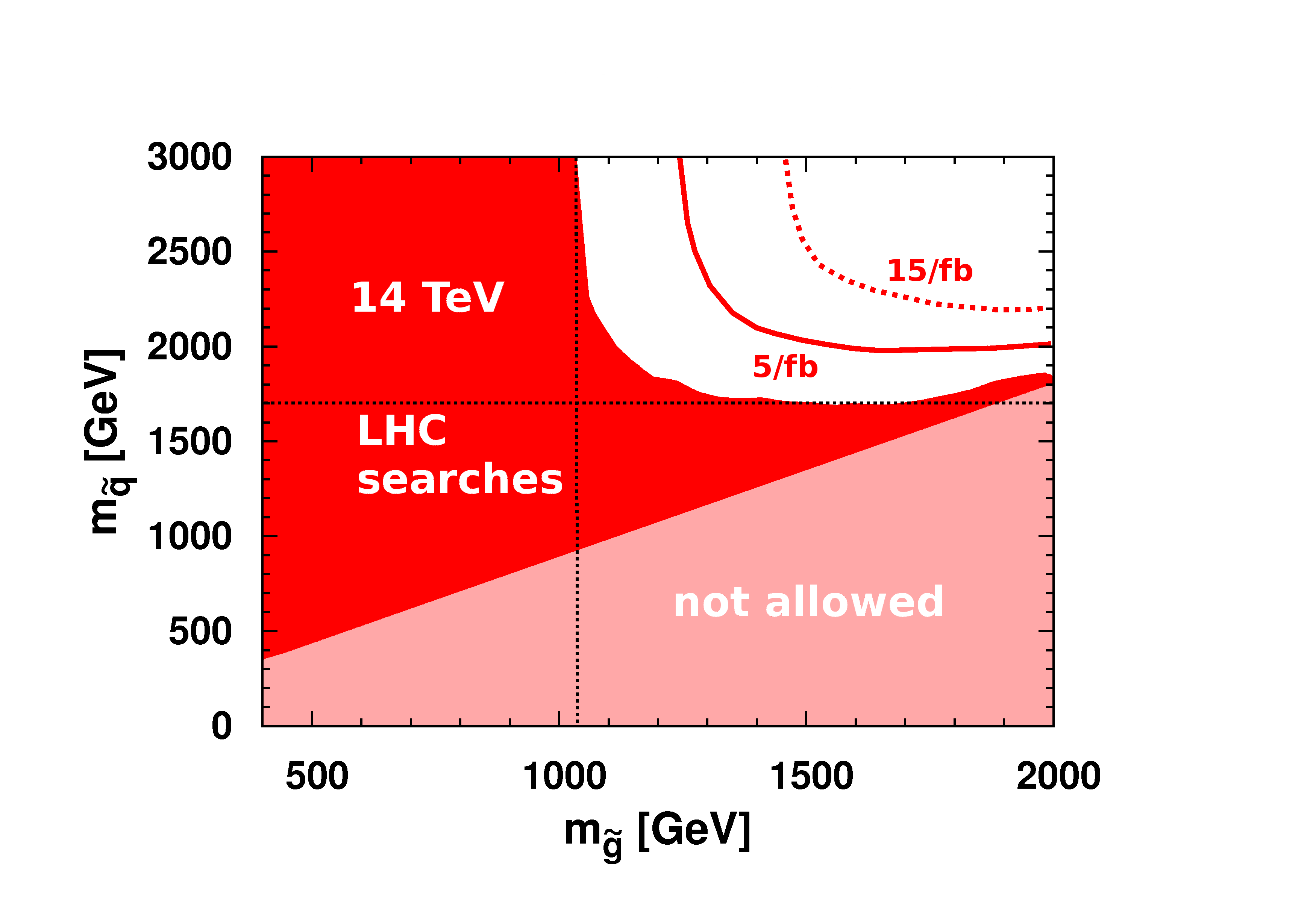}}
\vspace*{-5mm}
\end{center}
\caption{As in Fig.~\ref{f4}, but the excluded region is
translated into the $m_{\tilde q}, m_{\tilde g}$ plane. The
red area corresponds to the excluded regions for the integrated
luminosity slightly above 1~fb$^{-1}$; the expectations for 
higher luminosities have been indicated as well.}
\label{f5}
\end{figure}

The drop of the excluded region at large values of $m_0$ is due
to the fact that in this region the squarks become heavy. Here only 
higher energies will help, and doubling the LHC energy from 7 to
14~TeV, as planned in the coming years, quickly increases the
cross-section for the gluino production by orders of magnitude,
as shown in the right panel of Fig.~\ref{f4}. The expected
sensitivity at 14~TeV, plotted as the exclusion contour in case
nothing is found, assumes the same efficiency and luminosity
(slightly above one fb$^{-1}$ per experiment) as at 7~TeV.

These limits can be translated to the squark and gluino masses
and lead to the regions
indicated as not allowed ones in Fig.~\ref{f5}. Note that these
regions are forbidden in any model with the coupling between the
squarks and gluinos, so they are not specific to the CMSSM.
The squark masses below 1.1~TeV and the gluino masses below
0.62~TeV are excluded for the LHC data at 7~TeV, as shown in
the left panel of Fig.~\ref{f5}. The expected sensitivities for 
higher integrated luminosities at 7 and 14~TeV have been
indicated as well. One observes that increasing the energy is
much more effective than increasing the luminosity. At 14~TeV
the squarks with masses of 1.7~TeV and gluinos with masses of
1.02~TeV are within reach with 1~fb$^{-1}$ per experiment,
as shown in the right panel of Fig.~\ref{f5} \cite{bbkr2}.

\section{Indirect search for SUSY at the LHC}

Indirect manifestation of SUSY presumably takes place via virtual loops containing super partners. 
The most promising cases are the ones where the contribution of heavy super partners is enhanced by large $\tan\beta$.
One usually considers  the branchings of rare decays $B_s\to s\gamma, B_s\to \mu^+\mu^- , B_s\to \tau\nu $ and the $g-2$ factor of muon.
These quantities are relatively small and are measured with high precision (see Table below).

\begin{figure}[htb]
\begin{center}
\includegraphics[width=0.33\textwidth]{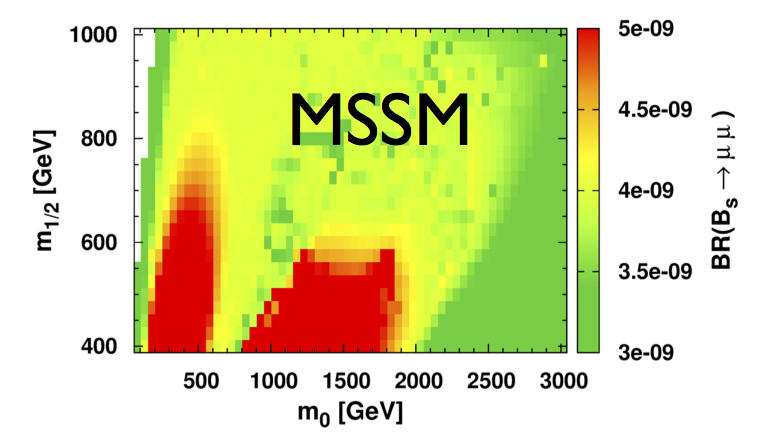}
\includegraphics[width=0.33\textwidth]{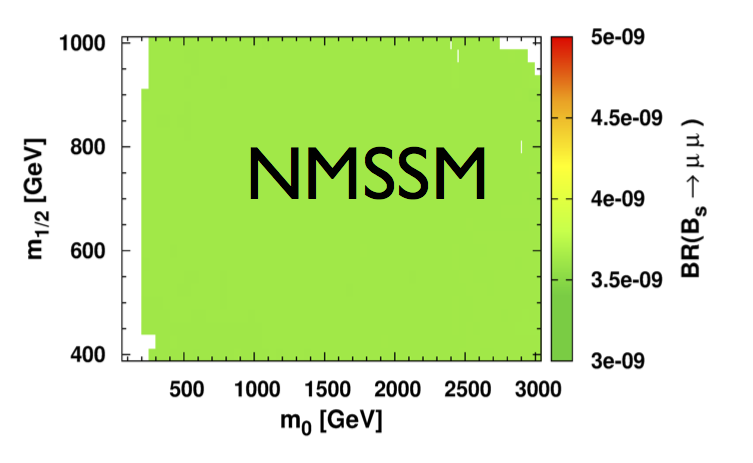}
\includegraphics[width=0.32\textwidth]{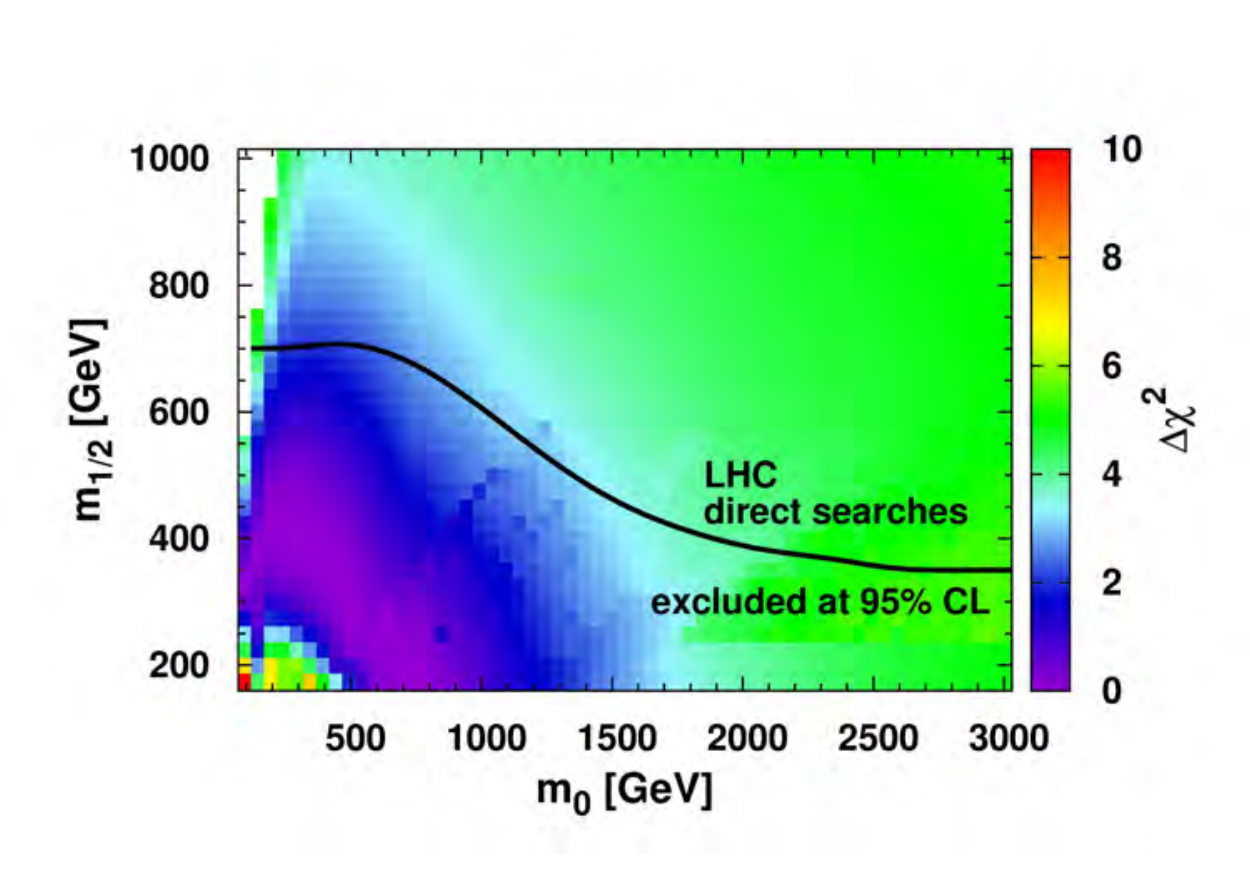}
\end{center}\vspace{-0.2cm}
\caption{Allowed region of the parameter space in the MSSM and the NMSSM with account of electroweak constraints. The right panel shows the region preferred by the $g-2$ constraint which is almost forbidden due to the LHC searches}
\label{f6}
\end{figure}

The most intriguing situation was with the $B_s\to \mu^+\mu^- $ decay. Its branching ratio is extremely small 
while in the MSSM one gets an enhancement $\sim \tan\beta^6$ that seems to contradict the measurement for relatively large $\tan\beta$. However, this is not exactly the case. The branching ratio in the MSSM is \cite{Ar}
\begin{eqnarray}
BR(B_s\to\! \mu^+\mu^-) = \frac{2\tau_BM_B^5}{64\pi} f^2_{B_s}
\sqrt{1\!-\!\frac{4m_l^2}{M_B^2}} 
\left[ ( \!1\!-\!{{4m_l^2}\over{M_B^2}}\! )
\Biggl| {{c_s\!-\!c_s'}\over{m_b\!+\!m_s}} \Biggr|^2\!+\!
\Biggl| {{c_p\!-\!c_p'}\over{m_b\!+\!m_s}}\!+\!
2{m_{\mu}\over M_{B}^2}(c_a\!-\!c_a') \Biggr|^2 \!
\right] 
\label{eq:bsmm_br_msugra}
\end{eqnarray}
  For large $\tan\beta$, the dominant
contribution to $c_s$ is given approximately by
\begin{equation}
c_s \simeq
{{G_F\alpha}\over {\sqrt 2\pi}}V_{tb}V_{ts}^*\!
\left( \frac{\tan^3\beta}{4\sin^2\theta_W} \right)\!
\left(\frac{m_b m_{\mu} m_t \mu}{M_W^2 M_A^2} \right)\!
\frac{\sin2\theta_{\tilde t}}{2}
\left(\! \frac{ m_{\tilde t_1}^2
\log\left( m_{\tilde t_1}^2 /\mu^2\right)}
{\mu^2-m_{\tilde t_1}^2} \!-\!
\frac{ m_{\tilde t_2}^2
\log\left({m_{\tilde t_2}^2 / \mu^2}\right)}
{\mu^2-m_{\tilde t_2}^2}\! \right)
\end{equation}
and being enhanced by $\tan\beta^3$ it is suppressed if the masses of stop1 and stop2 are degenerate. All together this gives the forbidden region in the parameter space shown in Fig.\ref{f6}.

Combination of constraints following from the rare decays are shown in Fig.\ref{f6} for the MSSM and NMSSM cases. All observables were calculated with the public code NMSSMTools \cite{NMSSM}.
We show also the requirement following from the $g-2$ factor confronted with the LHC data.

\section{Astrophysics search}

Search for manifestation of supersymmetry in astrophysics is based on the assumption that Dark matter  is composed of a super particle - the lightest neutralino. This is a natural consequence of R-parity preserving theories where the LSP is stable.  Minor R-parity violation with long-lived LSP is also possible. Within this scenario one has to provide the right amount of the DM abundance and possible manifestation of DM is supposed to take place in cosmic rays and in underground experiments.

\begin{figure}[htb]
\begin{center}\vspace{-0.1cm}
\includegraphics[width=0.80\textwidth]{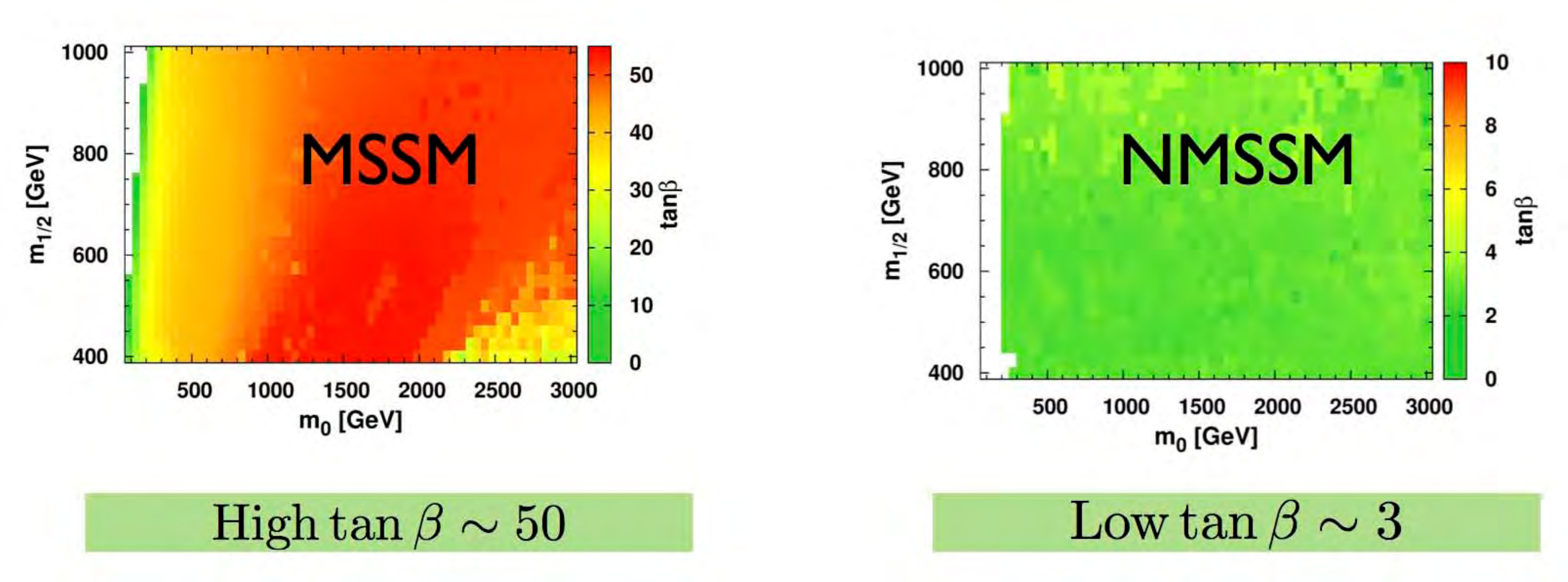}
\end{center}\vspace{-0.5cm}
\caption{Optimized values of $\tan\beta$ within the MSSM (left) and the NMSSM (right). One can see that high values of $\tan\beta$  are preferred within the MSSM except for the co annihilation regions, whereas low values are favored within the NMSSM}
\label{f9}
\end{figure}

Relic abundance of DM is given by the Boltzman equation \cite{Kolb}
\begin{equation}
\frac{dn_\chi}{dt}+3Hn_\chi=-<\sigma v>(n_\chi^2-n_{\chi,eq}^2), \ \ \ H=\dot{R}/R,
\end{equation}
where $<\sigma v>$ is the average annihilation cross-section.  From this equation one finds that
\begin{equation}
\Omega_\chi h^2=\frac{m_\chi n_\chi}{\rho_c}\approx\frac{3\cdot 10^{-27} cm^3 sec^{-1}}{<\sigma v>}
\end{equation}
For a given relic abundance $ \Omega_\chi h^2\approx 0.113\pm0.009$ \cite{Komatsu:2010fb} and $v\sim 300$ km/sec this gives
$$\sigma \sim 10^{-34} \ cm^2 =100\ pb$$
independently of the mass.
Fitting this value of the cross-section one gets the distribution of $\tan\beta$ values for the MSSM and NMSSM shown in Fig.\ref{f9} \cite{bbkr2}.

\begin{figure}[htb]
\begin{center}
\includegraphics[width=0.55\textwidth]{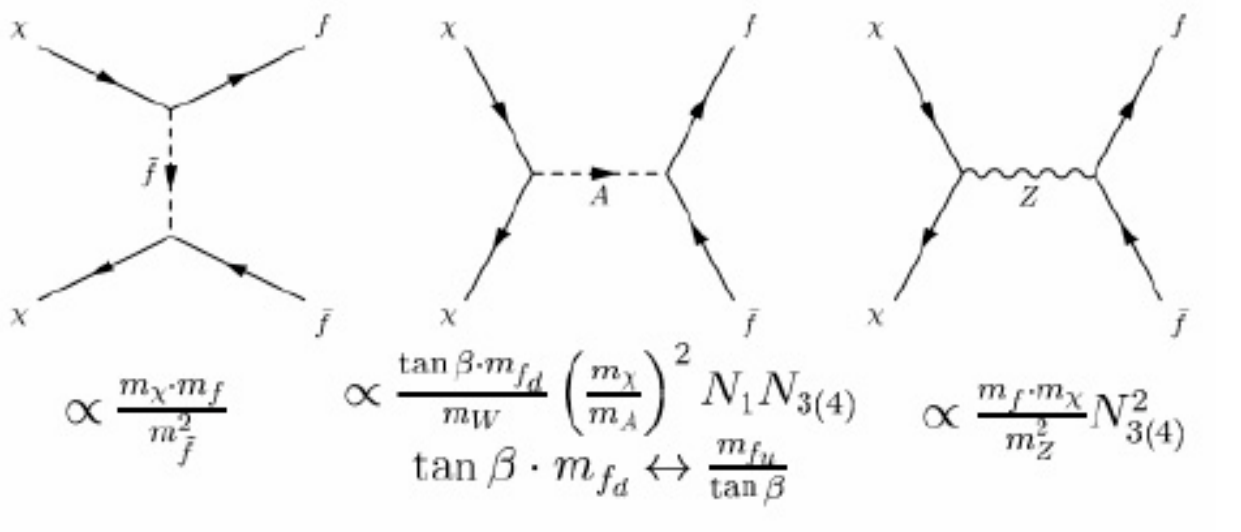}
\includegraphics[width=0.40\textwidth]{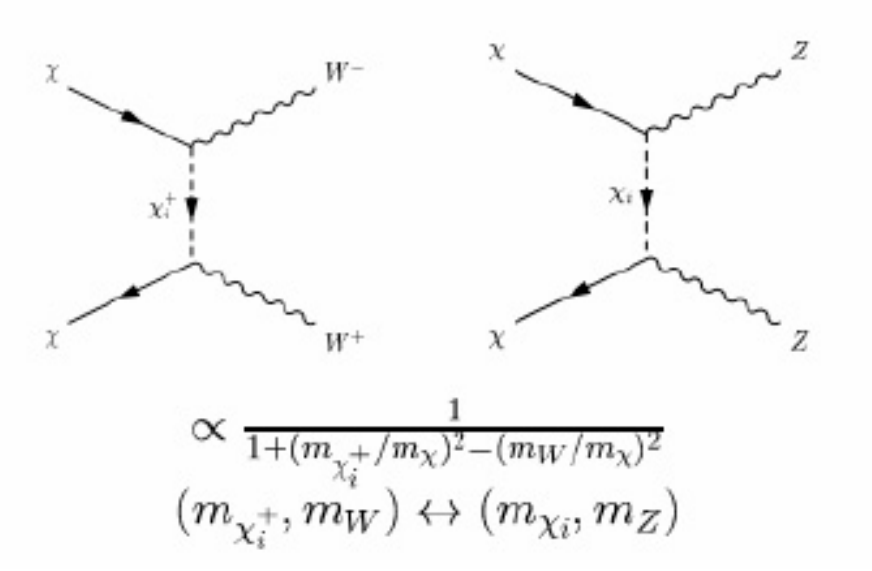}
\end{center}
\caption{DM annihilation diagrams in MSSM and NMSSM}
\label{f10}
\end{figure}

In the case when Dark matter has supersymmetric origin and consists of neutralino, one expects a signal in cosmic rays due to the annihilation of DM in the halo of the Galaxy. The annihilation diagrams are shown in Fig.\ref{f10}. This signal is rather weak, and noticeable enhancement in cosmic ray spectra is possible if  the density of  DM is boosted by clumpiness. The signal was searched  in positron, antiproton and diffuse gamma-ray spectra by EGRET, FERMI, PAMELA, PLANK and other telescopes but without definite success so far.

\begin{figure}[htb]
\begin{center}
\raisebox{35pt}{\includegraphics[width=0.49\textwidth]{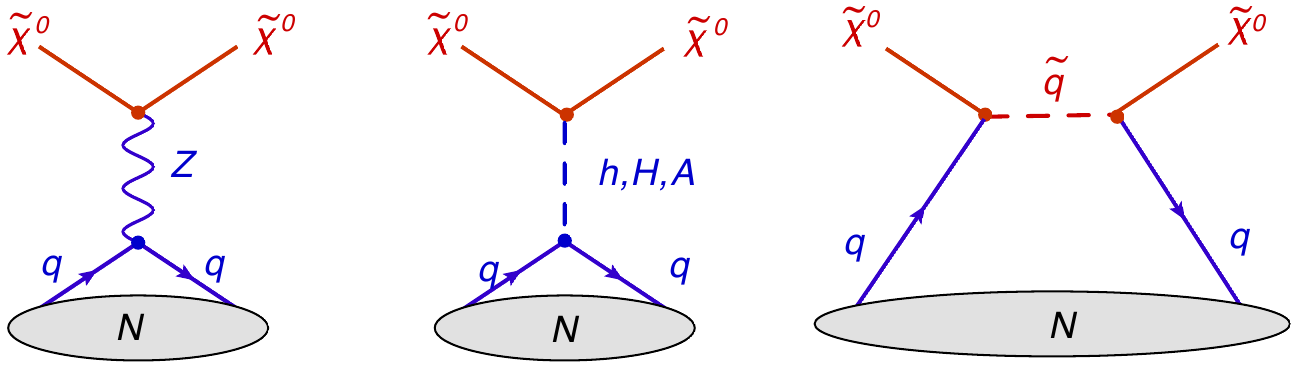}}
\includegraphics[width=0.40\textwidth]{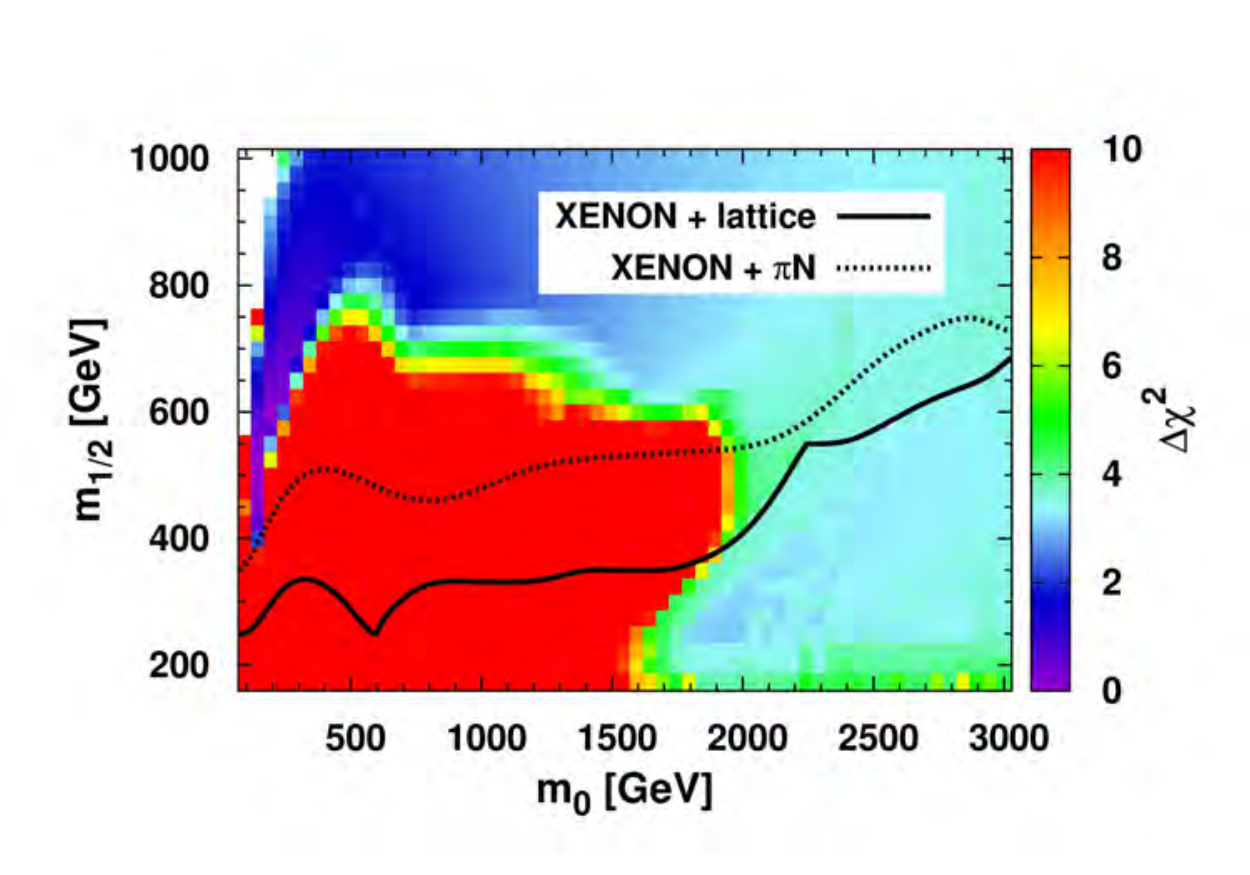}
\end{center}
\caption{The excluded region  from the XENON100 cross section limit for two values of the form factors. The red area is excluded from the EW constraints. }
\label{f7}
\end{figure}

In the underground experiments one can hope to observe interaction with  DM measuring the recoil energy. 
The cross section for direct scattering of WIMPS on nuclei has an experimental upper limit of about $10^{-8}$ pb, i.e., many orders of magnitude below the annihilation cross section. Scattering of the LSP on nuclei can only happen via elastic scattering, provided R-parity is conserved. The corresponding diagrams are shown in Fig.\ref{f7} (left). There are several experiments of this type: DAMA, Zeplin, CDMS, Edelweiss and more recently XENON100. Still, today we have no convincing evidence for direct dark matter detection. The excluded region in the $m_0-m_{1/2}$ plane from the XENON100 cross section limit \cite{Aprile:2011hi} is shown in Fig.\ref{f7} (right).

\begin{figure}[htb]
\begin{center}
\includegraphics[width=0.70\textwidth]{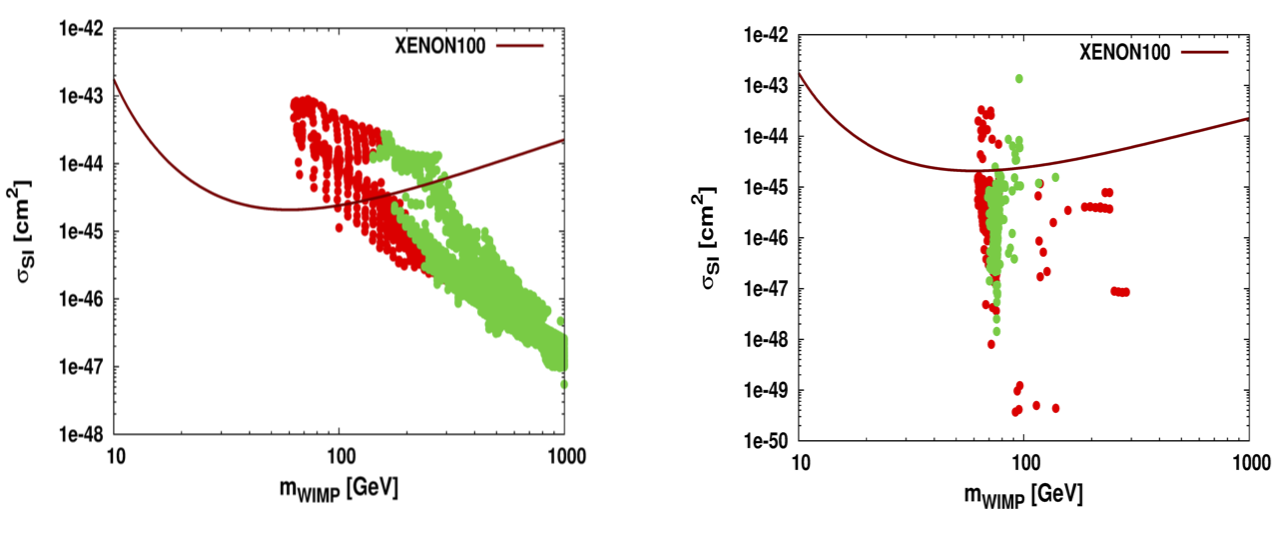}
\includegraphics[width=0.70\textwidth]{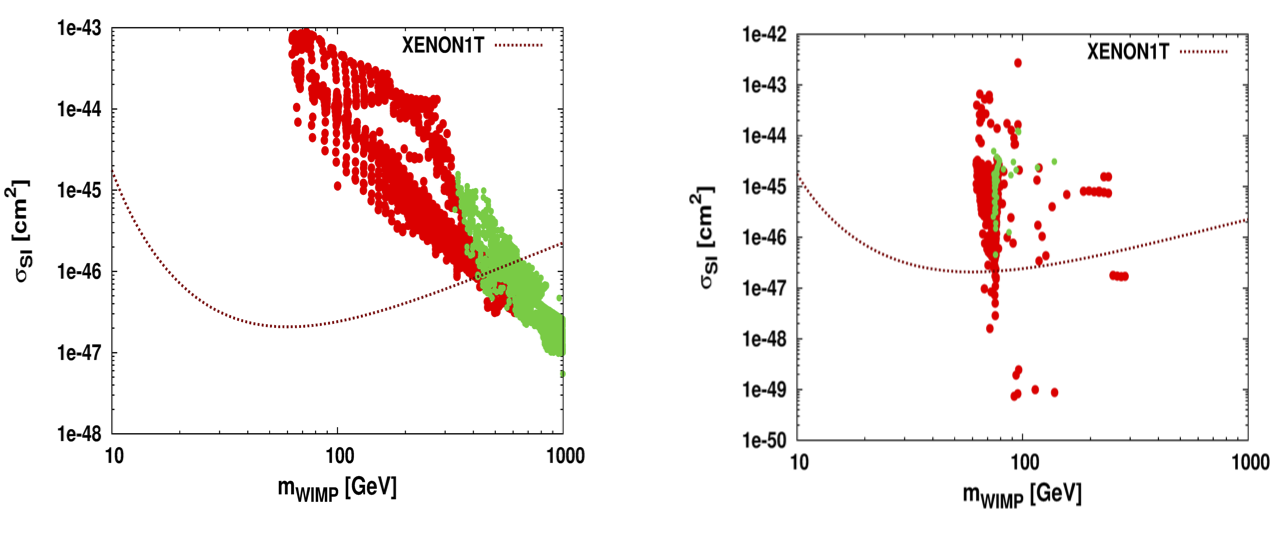}
\end{center}
\caption{The exclusion plot for direct interaction of WIMP DM with the nuclei. Left: MSSM, right: NMSSM.}
\label{f12}
\end{figure}

The typical exclusion plots for the cross-sections versus the WIMP mass plane  are shown in Fig.\ref{f12} in the case of the MSSM (left panel) and the NMSSM (right) panel. The area above the horizontal line is excluded by the XENON100 experiment (upper row) and by foreseen XENON1000 (lower row).  The  color area corresponds to SUSY fits. The red (dark) part is excluded by the LHC searches. One can see that in the MSSM the allowed WIMP mass is moving toward very high values while in the NMSSM the value around 100 GeV is preferable.

\section{Combined fit to all data}

\begin{figure}[htb]
\begin{center}
\includegraphics[width=0.33\textwidth,height=4.2cm]{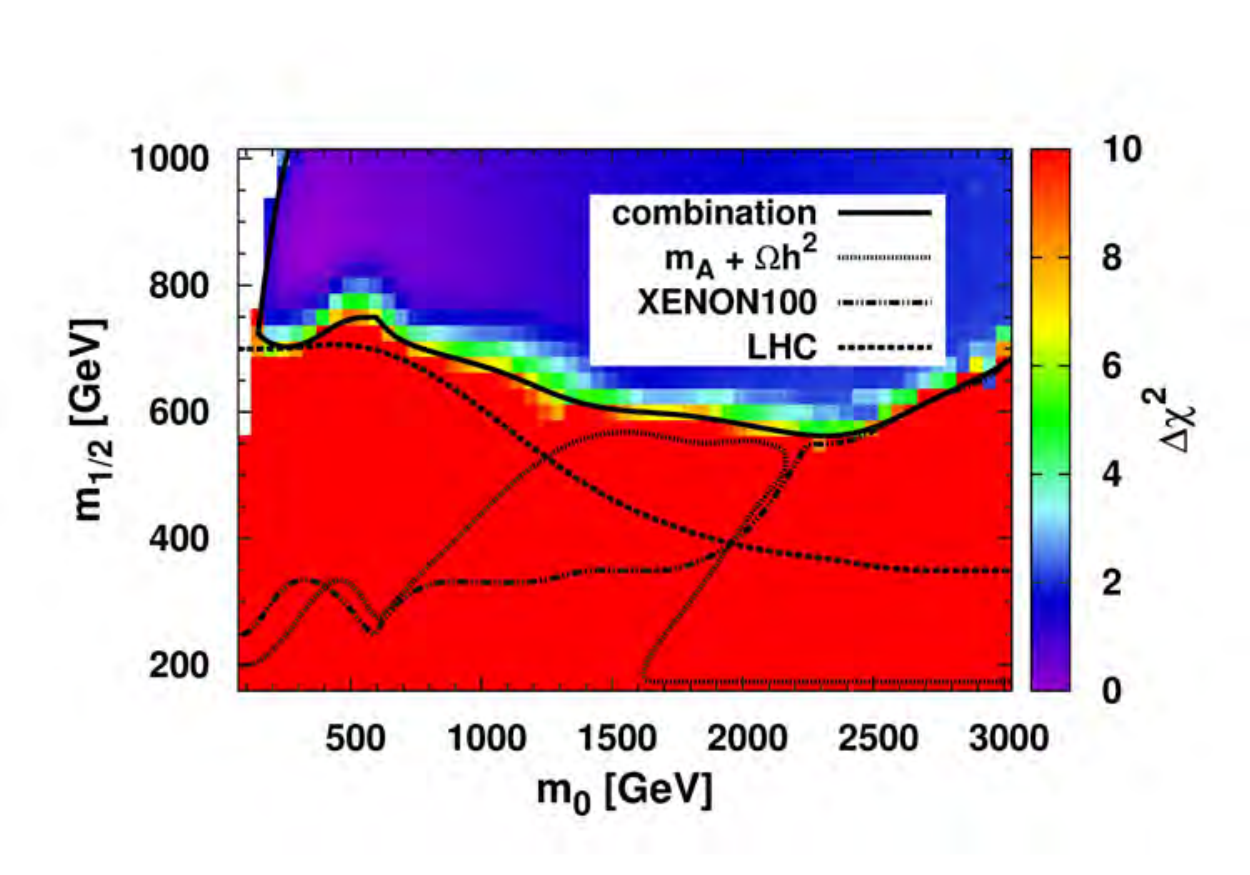}
\includegraphics[width=0.33\textwidth,height=4.2cm]{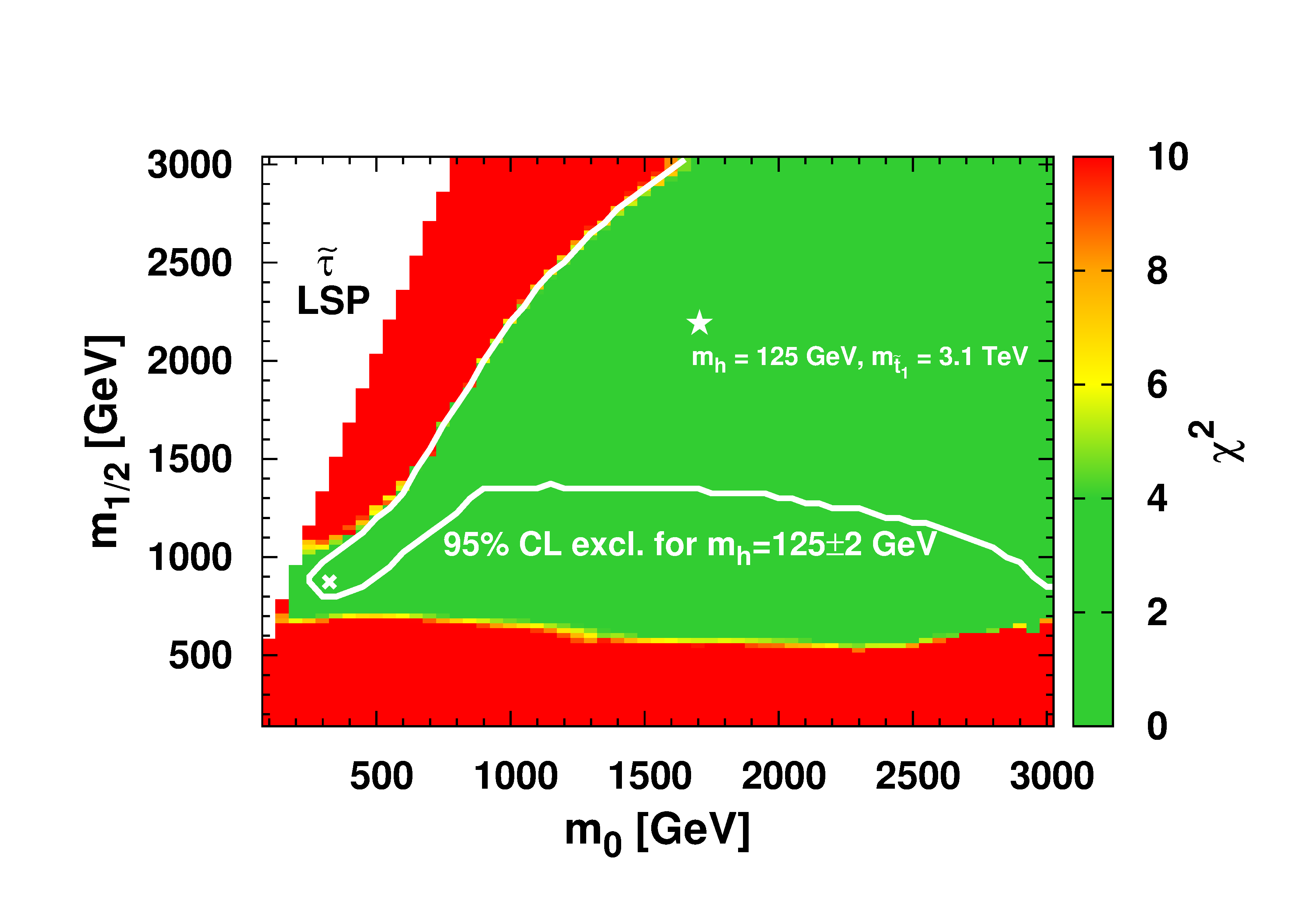}
\includegraphics[width=0.32\textwidth,height=4cm]{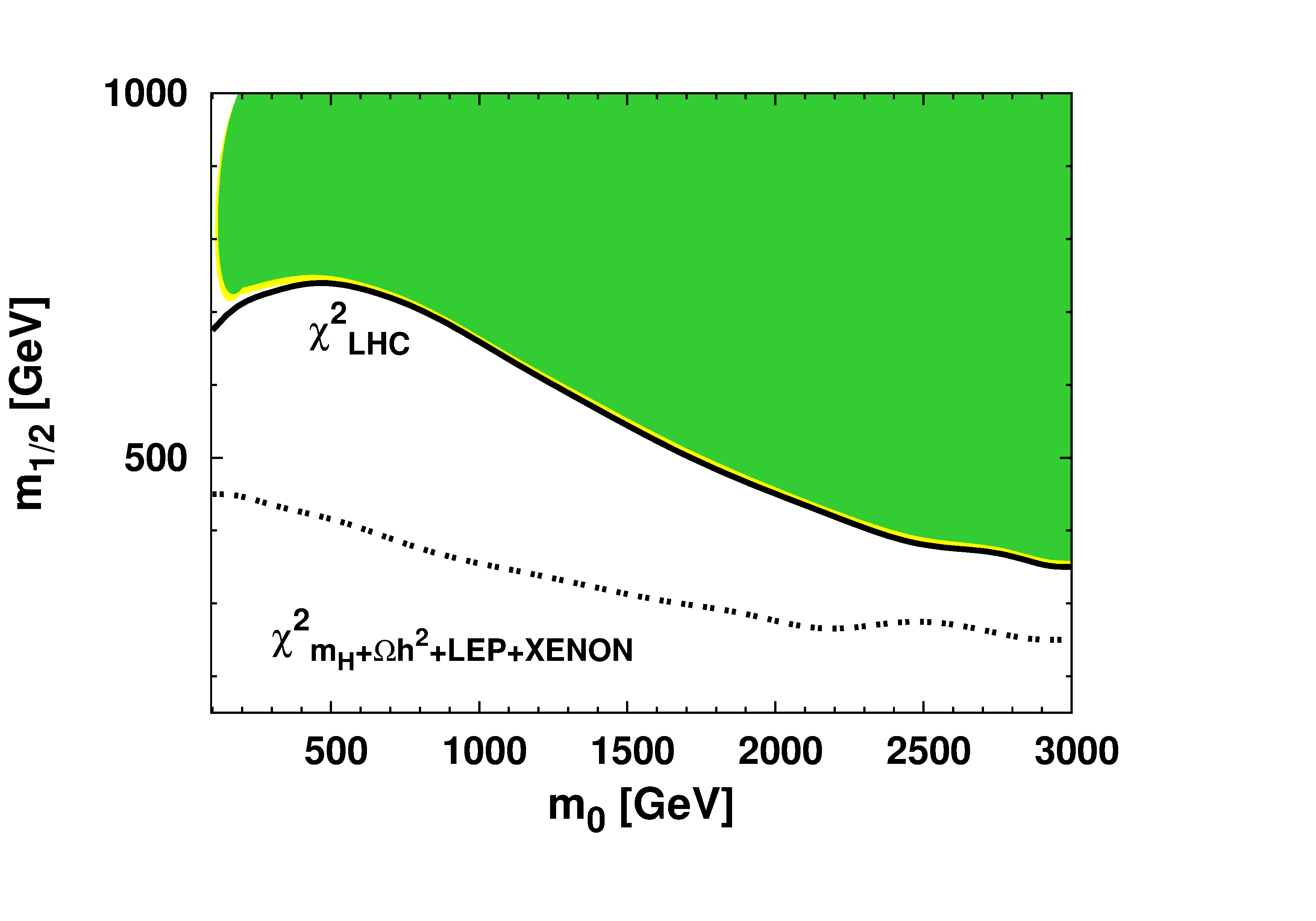}
\end{center}
\caption{Left: Combined constraints from the LHC searches, the relic density from WMAP, the direct DM searches from XENON100, limits on the pseudo-scalar Higgs mass and $g-2$ of muon (without the 125 GeV Higgs boson mass constraint) in the MSSM. Central: The account of the 125 GeV Higgs boson mass  with the 2 GeV mass uncertainty. The region below the white line is excluded at 95\% C.L.
Right: The same for the NMSSM.}
\label{f8}
\end{figure}

If one combines the excluded regions from the direct searches at the LHC, the relic density from the WMAP, the already stringent limits on the pseudo-scalar Higgs mass with the XENON100 limits, one obtains the excluded region shown in the left panel of Fig.\ref{f8} \cite{bbkr2}. The left panel corresponds to the MSSM.
The list of applied constraints is given in Table \ref{t1}.  Here the $g-2$ limit is included with the conservative linear addition of theoretical and experimental errors. One observes that the combination excludes $m_{1/2}$ below 525 GeV in the CMSSM for $m_0< $ 1500 GeV, which implies the lower limit on the WIMP mass of 230 GeV and a gluino mass of 1370 GeV, respectively.

If a Higgs mass of the lightest Higgs boson of 125 GeV is imposed, the preferred region is well above this excluded region, but the size of the preferred region is strongly dependent on the size of the assumed theoretical uncertainty. Accepting the 2 GeV uncertainty we get the excluded region shown in Fig.\ref{f8} (central panel), which is far above the existing LHC limits and leads to strongly interacting super partners above 2 TeV. However, in models with an extended Higgs sector, like the NMSSM, a Higgs mass of 125 GeV can be obtained for lower values of $m_{1/2}$, in which case the regions excluded in the MSSM become viable (see Fig.\ref{f8} right panel).

\begin{table}[htb]
\centering
\begin{tabular}{lll}
\hline\noalign{\smallskip}
Constraint & Data & Ref.  \\
\noalign{\smallskip}\hline\noalign{\smallskip}
$\Omega h^2$ & $0.113\pm 0.004$ & \cite{Komatsu:2010fb} \\
 $b\to s\gamma$ & $(3.55 \pm 0.24)\cdot 10^{-4}$ & \cite{hfag} \\
   $b\to\tau\nu$ &  $(1.68\pm 0.31)\cdot 10^{-4}$ & \cite{hfag} \\
    $\Delta a_\mu$ & $(302~\pm~63 (exp)~\pm~61 (theo))\cdot 10^{-11}$ &  \cite{Bennett:2006fi}\\
   $b_s\to\mu\mu$ &  $(3.2\pm 1.4)\cdot 10^{-9}$ & \cite{Aaij:2012ac}\\
$m_h$  & $ m_h > 114.4$ GeV & \cite{Schael:2006cr}\\
$m_A$ & $m_A > 480$ GeV for $\tan\beta \approx 50$& \cite{Chatrchyan:2012vp}\\
ATLAS & $ \sigma^{SUSY}_{had} < 0.003-0.03 $ pb & \cite{ATLAS-CONF-2012-033}\\
CMS & $\sigma^{SUSY}_{had} < 0.005-0.03 $ pb &\cite{CMS-PAS-SUS-12-005} \\
XENON100 & $\sigma_{\chi N} < 8 \cdot 10^{-45}-2\cdot 10^{-44} cm^2$& \cite{Aprile:2011hi}\\
\noalign{\smallskip}\hline
\end{tabular}
\caption{List of all constraints used in the fit to determine the excluded region of the CMSSM parameter space. }
\label{t1}
\end{table}
\section{Conclusion}
The SUSY phenomenology is a wide and well explored subject today. It seems that from the experimental point of view we are approaching crucial times. Today SUSY is far from being dead. The situation can be summarized as follows:

$\bullet$ There is no signal so far, but do not give up

$\bullet$ There is still plenty of room for SUSY

$\bullet$ Interpretations of searches are model dependent

$\bullet$  The LHC run at 14 TeV might be crucial for low energy SUSY

Due to remarkable properties, SUSY has many followers. It has almost no competitors. One may exclaim: Give me something better and I will stick to it but so far I like SYSY!



\section*{References}
\begin{thebibliography}{99}
\bibitem{ATLAS_SUSY}
ATLAS collaboration:
ATLAS-CONF-2012-145,
%
ATLAS-CONF-2012-151,
ATLAS-CONF-2012-103,
%
ATLAS-CONF-2012-105.
\bibitem{CMS_SUSY_pub}
CMS Collaboration:
CMS-PAS-SUS-11-016,CMS-SUS-11-024.

\bibitem{CMS_SUSY_web}
\texttt{https://twiki.cern.ch/twiki/bin/view/CMSPublic/SUSYSMSSummaryplots}
\bibitem{ATLAS_stop1}
ATLAS collaboration:
\texttt{arXiv:1208.4305},\texttt{arXiv:1209.2102}


\bibitem{bbkr2}
C. Beskidt, W. de Boer, D.I. Kazakov, F. Ratnikov,
\emph{Eur. Phys. J.} \textbf{C72} (2012) 2166;
\emph{JHEP} \textbf{1205} (2012) 094; NMSSM fits to appear.
\bibitem{Ar} R.L. Arnowitt, B. Dutta, T. Kamon, and M. Tanaka,
{\em Phys. Lett.} {\bf B538} (2002) 121.
\bibitem{NMSSM} D. Das, U. Ellwanger, and A. M. Teixeira, "NMSDECAY: A Fortran Code for SUSY Particle Decays in the NMSSM", Comput.Phys.Commun. 183 (2012) 774–779, arXiv:1106.5633.
\bibitem{Kolb} E. Kolb and M.S. Turner, {\em The Early Universe, Frontiers in Physics}, Addison Wesley, 1990.
\bibitem{Komatsu:2010fb}
E. Komatsu et~al.,
\emph{Astrophys. J. Suppl.} \textbf{192} (2011) 18.

\bibitem{hfag}
\texttt{http://www.slac.stanford.edu/xorg/hfag/rare/ichep10/radll/OUTPUT/TABLES/}
\texttt{radll.pdf}

\bibitem{Bennett:2006fi}
Muon G-2 Collaboration,
\textit{Phys. Rev.} \textbf{D73} (2006) 072003.

\bibitem{Aaij:2012ac}
LHCb collaboration,
\textit{Phys. Rev. Lett.} \textbf{108} (2012) 231801

\bibitem{Schael:2006cr}
ALEPH Collaboration, DELPHI Collaboration, L3 Collaboration,
OPAL Collaborations, LEP WG for Higgs Boson Searches
Collaboration, \textit{Eur. Phys. J.} \textbf{C47} (2006) 547.

\bibitem{Chatrchyan:2012vp}
CMS Collaboration, \textit{Phys. Lett.} \textbf{B713} (2012) 68.
ATLAS Collaboration, \emph{Phys. Lett.} \textbf{B705} (2011) 174.

\bibitem{ATLAS-CONF-2012-033}
ATLAS Collaboration:
ATLAS-CONF-2012-033.

\bibitem{CMS-PAS-SUS-12-005}
CMS Collaboration: 
CMS-PAS-SUS-12-005.

\bibitem{Aprile:2011hi}
E. Aprile, et~al.,
\textit{Phys. Rev. Lett.} \textbf{107} (2011) 131302.

\end{thebibliography}
\end{document}